\documentclass[fleqn,usenatbib]{mnras}

\usepackage{newtxtext,newtxmath}

\usepackage[T1]{fontenc}

\DeclareRobustCommand{\VAN}[3]{#2}
\let\VANthebibliography\thebibliography
\def\thebibliography{\DeclareRobustCommand{\VAN}[3]{##3}\VANthebibliography}
\newcommand\rmd{{\rm d}}

\usepackage{graphicx}	
\usepackage{amsmath}	
\usepackage{orcidlink}
\usepackage{widetext}
\usepackage{cuted}

\title[Synchrotron spectropolarimetry]{Spectropolarimetry of synchrotron radiation  
  from relativistic electrons \\ 
  with anisotropic pitch-angle and 
  various energy distributions }

\author[P.C.W. Lai et al.]{
Paul C. W.~Lai $^{\rm \orcidlink{0000-0003-3601-5127}}$,$^{1}$\thanks{E-mail: chong.lai.22@ucl.ac.uk (PCWL); kinwah.wu@ucl.ac.uk (KW); yapyeexuan@gapp.nthu.edu.tw (YXJY)}
Kaye J.~Li $^{\rm \orcidlink{0000-0002-1657-0265}}$,$^{1}$
Y. X.~Jane Yap $^{\rm \orcidlink{0000-0002-3349-3089}}$,$^{2, 1\star}$
Kinwah Wu $^{\rm \orcidlink{0000-0002-7568-8765}}$ $^{1,3,4,5\star}$
and Albert K. H.~Kong $^{\rm \orcidlink{0000-0002-5105-344X}}$ $^{2}$
\\
$^{1}$Mullard Space Science Laboratory, 
University College London,  
Holmbury St Mary, Surrey RH5 6NT, United Kingdom\\
$^{2}$Institute of Astronomy, 
National Tsing Hua University,  
 Hsinchu 30013, Taiwan (R.O.C.)\\
$^{3}$Research School of Astronomy and 
  Astrophysics, Australian National University 
Canberra ACT 2611, Australia\\
$^{4}$Department of Physics, 
  Chinese University of Hong Kong, 
  Shatin, NT, Hong Kong SAR, China\\
  $^{5}$Kavli Institute for the Physics and Mathematics of the Universe (WPI), UTIAS,  
The University of Tokyo, Kashiwa, Chiba 277-8583, Japan
}

\date{Accepted XXX. Received YYY; in original form ZZZ}

\pubyear{\the\year{}}

\begin{document}
\label{firstpage}
\pagerange{\pageref{firstpage}--\pageref{lastpage}}
\maketitle

\begin{abstract}
For synchrotron radiation from relativistic electrons having a power-law energy distribution   with a power-law index $p$ in an optically thin medium with a locally uniform magnetic field, it is generally adopted that the spectral index $\alpha = (p-1)/2$  and the degree of linear polarisation  $\Pi_{\rm L,pl} = (p +1)/(p+7/3)$, and hence that $\Pi_{\rm L,pl} = (\alpha +1)/(\alpha+5/3)$.
These ($\alpha$, $\Pi_{\rm L,pl}$\;\!; $p$) relations  
  are derived assuming that 
  the electrons have an isotropic 
  momentum distribution 
  and a power-law energy distribution, 
  and they have been used, 
  almost universally, in the interpretation 
  of polarimetry observations. 
In this study, we assess 
  the validity 
  of the ($\alpha$, $\Pi_{\rm L,pl}$\;\!; $p$) 
  relations in different scenarios,
  such as anisotropic electron momenta
  and non-power-law distributions of electron energy.
We calculate the synchrotron radiation 
  and polarisation spectra 
  for power-law, kappa 
  and log-parabola 
  electron-energy distributions,  
  with isotropic and two anisotropic 
  (beamed and loss cone) 
  electron-momentum distributions. 
Our calculations 
  show that when the electron momenta are isotropic, 
  the usual ($\alpha$, $\Pi_{\rm L,pl}$\;\!; $p$) relations are generally applicable. 
However, if the electrons are anisotropic, 
  the usual ($\alpha$, $\Pi_{\rm L,pl}$\;\!; $p$) relations could break down. 
Applying the  
  usual ($\alpha$, $\Pi_{\rm L,pl}$\;\!; $p$) relations  
  indiscriminately
  without caution on anisotropy in 
  the electron momentum distribution,  
  would lead to incorrect interpretations of polarimetry data. 
\end{abstract}




%
\section{Introduction}  
\label{sec:intro}

Synchrotron radiation,  
  emitted when relativistic charged 
  particles gyrating around a magnetic field, 
  is a common radiative process 
  in astrophysical environments.  
A characteristic of synchrotron radiation 
  is its strong linear polarisation 
  but negligible circular polarisation. 
The spectrum of synchrotron radiation  
  from an ensemble of relativistic electrons 
  with an isotropic 
  extended power-law energy distribution, 
  ${\rm d}n_{\rm e}(\gamma)/{\rm d}\gamma \propto \gamma^{-p}$  
  (where $n_e(\gamma)$ 
  is the electron number density\footnote{ 
    The number density of electrons 
       with energy in the range $(\gamma_1, \gamma_2)$,   
         at a given location,  is 
    \begin{align} n_{\rm e} = \int_{\gamma_1}^{\gamma_2} 
         {\rm d}\gamma \ 
         \left(\frac{{\rm d}n_{\rm e}(\gamma)}{{\rm d}\gamma}\right) \ . \nonumber 
         \end{align} 
Note that in some literature, 
      the expression  ${{\rm d}n_{\rm e}(\gamma)}/{{\rm d}\gamma}$ 
      that we adopt here is 
        referred to as ``$n_{\rm e}(\gamma)$''.}, 
         $\gamma$ is the Lorentz factor of the electrons), 
         of energy index $p$  
     is also a power law, in the optical thin regime, 
     and the specific intensity, 
      at frequency $\nu$, is given by 
      $I_\nu \propto \nu^{-\alpha}$. 
     where $\alpha = (p-1)/2$ 
     \citep[see e.g.][]{Tucker1975rpa,Rybicki1979rpa}. 
In the optically thick regime,
    the emission spectrum also follows a power law but with a different spectral index of $\alpha = -5/2$, 
   which is independent of 
   the electron energy distribution 
   \citep{Ginzburg1965ARA&A,Ginzburg1969ARA&A}.  
The degree of linear polarisation or
the linear polarisation degree (polarisation degree or PD hereafter) is
\begin{equation} 
\label{eq:pd_pl}
    \Pi_{\rm L, pl} 
    = \frac{p+1}{p+7/3}
    = \frac{\alpha+1}{\alpha+5/3}
\end{equation}
    for optically thin emission 
  and $\Pi_{\rm L,pl}^{\rm thick} = 3/(6p +13)$ for optically thick emission 
   \citep{Ginzburg1969ARA&A}.   
These expressions  
  are derived with the 
  assumption of a uniform magnetic field  
  orderly aligned in the emission region,     
   and the polarisation vector 
   is thus perpendicular 
   to the orientation of the magnetic field 
   ${\boldsymbol B}$.  
They are applicable 
   to other charged leptons and charged baryons besides electrons.
These expressions have been commonly used in astronomy,  
  for the interpretation of the polarised radiation 
  from a variety of non-thermal sources, 
  including jets from active galactic nuclei (AGN), 
  X-ray binaries, gamma-ray bursts
  and diffuse emission from extended systems 
  such as supernova remnants
  and pulsar-wind nebulae
    \citep[e.g.,][]{Coburn2003Natur, Perlman2011ApJ, Lai2022ApJ, Kaaret2024ApJ, Slane2024Galax}.

Power-law-like energy distributions of charged particles 
  can be generated from Fermi I process 
  (e.g., in diffusive shocks) 
  and Fermi II process (e.g., 
   in randomly moving deflectors/reflectors).
A perfect power-law energy distribution 
   for the emitting charged particles 
   is, however, an idealised situation. 
Radiative cooling 
  and processes such as episodic injections of high-energy particles  
  would modify the particle energy distribution, 
  even if the energy distribution is a perfect 
  power law initially. 
Nonetheless, 
  for the analysis at fixed wavelength-band observations,  
  using a power-law distribution for computing 
  the emission spectrum and polarisation  
  is still a reasonable approximation 
  in most situations 
  (which we will demonstrate in the later sections). 

Note that the emitting charged particles were assumed 
  to have an isotropic momentum pitch-angle distribution (hereafter pitch-angle distribution),   
  with respect to the orientation of the magnetic field,  
  in the rest frame of the infinitesimal volume 
  containing the particles 
  when deriving the above expressions. 
This assumption provides a great simplification in 
  deriving the angle-dependent polarisation components  
  of the radiation -- 
  the emission, absorption and 
  other radiative transfer coefficients\footnote{With 
  the emission, absorption 
  and Faraday transfer coefficients   
  defined in the local rest frame 
  of the infinitesimal emission volume, 
  one can execute the radiative transfer calculations 
  using ray-tracing algorithms, in a covariant manner 
  \citep[see][]{Fuerst2004A&A,Younsi2012A&A}
  for systems with large 
  differential special relativistic 
  motion \citep[see Appendix A in][]{Saxton2010MNRAS}, 
  under extreme gravity 
  \citep[see e.g.,][]{Prather2023ApJ} 
  or in cosmological expansion 
  \citep[see][]{Chan2019MNRAS,On2019MNRAS}.}.  
With the advancement 
  in the broad multi-band polarimetric observations  
  and the new polarisation data obtained in high-energy domains, 
  e.g., current observations in the keV X-rays 
  by the Imaging X-ray Polarimetry Explorer 
  \citep[IXPE, see][]{Soffitta2021AJ,Weisskopf2022JATIS}, 
  it is timely to revisit 
  how the polarisation in the synchrotron radiation changes
  when relativistic charged particles 
  do not have an isotropic pitch-angle distribution 
  and do not have a power-law energy distribution.  
  Recent studies 
\citep{Yang2018ApJ, Comisso2020ApJ}
has demonstrated that we cannot directly infer the electron energy spectral index $p$ from the observed synchrotron spectral index $\alpha$ if the emitting electron momenta 
  do not have an isotropic distribution.
We extend the work of \cite{Yang2018ApJ} 
  by calculating the polarisations of the synchrotron radiation
  from a variety of anisotropic electron
  momentum distributions
  that are relevant to astrophysical applications.
We organise the paper as follows.  
We present in Section 2 
  the derivations 
  of the emissivities and polarisation 
  for synchrotron radiation 
  from relativistic charged particles,   
  relaxing the usual assumptions  
  of an extended power-law energy distributions  
  and of isotropic pitch-angle distributions, 
  and we show in Section 3 the results. 
We discuss the astrophysical implications in Section 4.   
A summary is given in Section 5.

%
\section{Synchrotron radiation from relativistic electrons} 
\label{sec:synchrotron}

%
\subsection{Radiative power and polarisation} 
\label{subsec:basic} 

Synchrotron radiation is anisotropic, 
  with the observed intensity 
  depending on the line-of-sight to the observer 
  with respect to the orientation of 
  the magnetic field and travelling direction of the charged particles in the emission region. 
Suppose the radiation propagating 
   in the direction ${\boldsymbol {\hat n}}$ 
   along the line of sight 
   and the magnetic field ${\boldsymbol B}$ 
   is locally uniform 
   and aligned with the $z$-axis.  
We can then decompose 
   the power of synchrotron radiation 
   into two orthogonal components, 
   designated as 
    the perpendicular component 
    and the parallel component.   
The perpendicular component corresponds to 
   the electromagnetic waves 
   whose electric field unit vector 
       ${\boldsymbol {\hat \epsilon}}_{\perp}$ 
       perpendicular to the magnetic field 
       ${\boldsymbol B}$ and 
       ${\boldsymbol {\hat n}}$, 
       and the parallel component 
       to the electromagnetic waves  
          with the electric field 
          unit vector given by 
  ${\boldsymbol {\hat \epsilon}}_{\parallel} 
    = {\boldsymbol {\hat n}}\times 
    {\boldsymbol {\hat \epsilon}}_{\perp}$.  
The power of the two components 
  of synchrotron radiation 
  from an electron 
  with a Lorentz factor $\gamma$  
  are\footnote{As discussed in \cite{Rybicki1979rpa}, 
    there would be an additional factor of 
    $1/\sin^2\tilde{\alpha}$ for the received power 
    if electrons move in perfect
    helical motion around a magnetic field line.
In astrophysical environments,  
   magnetic fields are hardly perfectly uniform and ordered. 
Electrons would move in curved helical orbits   
   along the magnetic field line, 
   and they can bounced back and forth by magnetic mirroring effect.
As such we should expect the received power and the emitted power 
  to match each other. 
An additional factor of $1/\sin^2\tilde{\alpha}$ 
    is therefore unnecessary 
    in the astrophysical applications.}   
\begin{align}
    \frac{\rmd W_\perp}{\rmd \omega \, \rmd \Omega} 
    (\omega, \tilde{\theta}, \tilde{\alpha}, \gamma) &=
    \left( \frac{\omega_B}{2\pi} \right)
    \frac{e^2}{3\pi^2 c}
    \left( \frac{\omega \rho}{c} \right)^2
    \left( \frac{1}{\gamma^2} + \tilde{\theta}^2 \right)^2
    [K_{2/3}(\xi)]^2 \label{eq:W_perp} \ , \\
    \frac{\rmd W_\parallel}{\rmd \omega \, \rmd \Omega} 
    (\omega, \tilde{\theta}, \tilde{\alpha}, \gamma) &=
    \left( \frac{\omega_B}{2\pi} \right)
    \frac{e^2}{3\pi^2 c}
    \left( \frac{\omega \rho}{c} \right)^2
    \tilde{\theta}^2
    \left( \frac{1}{\gamma^2} + \tilde{\theta}^2 \right)
    [K_{1/3}(\xi)]^2 \label{eq:W_para} \  
\end{align}
 \citep[see e.g.][]{Rybicki1979rpa, Yang2018ApJ}, 
  where 
    $\omega$ is the angular frequency of the radiation,
    $\tilde{\alpha}$ is the angle 
     between the electron momentum and the magnetic field (the pitch angle),
    $\tilde{\theta}$ is the angle between ${\boldsymbol {\hat n}}$ and 
    the instantaneous-trajectory plane,
    $\omega_B = eB/(\gamma m_e c)$ is the gyro frequency,
    $\rho = c/(\omega_B \sin\tilde{\alpha})$ is the radius of the electron's helical trajectory,
    $K_a$ is the modified Bessel function of the second kind with order $a$, 
    $\xi = (1/\gamma^2 + \tilde{\theta}^2)^{3/2} (\omega \rho)/(3c)$, and $B = |\;\!{\boldsymbol B}\;\!|$.  
The total power of the radiation is given by 
\begin{align}
    \frac{\rmd W_{\rm tot}}{\rmd \omega \, \rmd \Omega} 
    (\omega, \tilde{\theta}, \tilde{\alpha}, \gamma) &= 
    \frac{\rmd W_\perp}{\rmd \omega \, \rmd \Omega} + 
    \frac{\rmd W_\parallel}{\rmd \omega \, \rmd \Omega}  
    \ , 
\label{eq:W1}
\end{align}     
  and the power of the linearly polarised radiation by 
\begin{align} 
    \frac{\rmd W_{\rm pol}}{\rmd \omega \, \rmd \Omega} 
    (\omega, \tilde{\theta}, \tilde{\alpha}, \gamma) &= 
    \frac{\rmd W_\perp}{\rmd \omega \, \rmd \Omega} - 
    \frac{\rmd W_\parallel}{\rmd \omega \, \rmd \Omega} \ . 
\label{eq:W2} 
\end{align} 
As the perpendicular component 
  is larger than the parallel component,  
  the linear polarisation of synchrotron radiation 
  is always perpendicular to the magnetic field direction 
  at the location of emission.

The total power of the emission from an ensemble of electrons 
  is the sum of the radiative power of individual electrons 
  (for negligible self-absorption).  
We denote 
  the electron number distribution function
    as $N_e(\gamma, \tilde{\alpha})$.
The specific flux of synchrotron radiation  
  from an ensemble of electrons, 
  measured by an observer located at distance $D$, is \footnote{The subscript 
    $\omega$ here represents 
    per unit angular frequency $\omega$, i.e. 
    the specific flux $F_\omega = \rmd F/ \rmd \omega$, 
    where $F$ is the flux of the radiation integrated 
    over the relevant range of $\omega$  
    (cf. $\rmd W/\rmd \omega$ 
    in equation~\ref{eq:W1} and \ref{eq:W2}).}
\begin{align}
    F_\omega(\omega, \theta) = &
    \frac{2\pi }{D^2}
    \int_0^\pi \rmd \tilde{\alpha} \ 
       \sin\tilde{\alpha} \nonumber \\
    & \bigg[
    \int_{\gamma_{\rm min}}^{\gamma_{\rm max}}
    \rmd \gamma \ 
    \frac{\rmd W_{\rm tot}}{\rmd \omega \, \rmd \Omega}
    (\omega, \tilde{\alpha}-\theta, \tilde{\alpha}, \gamma)
    N_e(\gamma, \tilde{\alpha}) \bigg] \nonumber  \\
     = &
    \frac{2\pi}{D^2}
    \int_{-\theta}^{\pi-\theta}  \rmd \tilde{\theta} \ 
     \sin(\tilde{\theta}+\theta) \nonumber \\
    & \bigg[
    \int_{\gamma_{\rm min}}^{\gamma_{\rm max}}
     \rmd \gamma \ 
    \frac{\rmd W_{\rm tot}}{\rmd \omega \, \rmd \Omega}
    (\omega, \tilde{\theta}, \tilde{\theta}+\theta, \gamma)
    N_e(\gamma, \tilde{\theta}+\theta) \bigg] \ ,   
  \label{eq:F_omega}     
\end{align}
  and the corresponding polarised specific flux is  
\begin{align}
    F_{\omega,{\rm pol}}(\omega, \theta) = &
    \frac{2\pi }{D^2}
    \int_0^\pi \rmd \tilde{\alpha} \     \sin\tilde{\alpha} \nonumber \\
    & \bigg[
    \int_{\gamma_{\rm min}}^{\gamma_{\rm max}}
     \rmd \gamma \ 
    \frac{\rmd W_{\rm pol}}{\rmd \omega \, \rmd \Omega}
    (\omega, \tilde{\alpha}-\theta, \tilde{\alpha}, \gamma)
    N_e(\gamma, \tilde{\alpha}) 
    \bigg] \nonumber \\
    = &
    \frac{2\pi }{D^2}
    \int_{-\theta}^{\pi-\theta}  
     \rmd \tilde{\theta} \ 
     \sin(\tilde{\theta}+\theta) \nonumber \\
     & \bigg[
    \int_{\gamma_{\rm min}}^{\gamma_{\rm max}}
    \rmd \gamma \ 
    \frac{\rmd W_{\rm pol}}{\rmd \omega \, \rmd \Omega}
    (\omega, \tilde{\theta}, \tilde{\theta}+\theta, \gamma)
    N_e(\gamma, \tilde{\theta}+\theta) \bigg] \ , 
 \label{eq:F_pol_omega}
\end{align}  
  where $\theta$ is the angle between 
    the line-of-sight 
    directional unit vector ${\boldsymbol {\hat n}}$ to the observer 
    and 
    the magnetic field ${\boldsymbol {B}}$.  

If the electrons have an isotropic 
 pitch-angle  
  distribution with respect to the magnetic field, i.e., $N_e(\gamma, \tilde{\alpha})$ does not depend on $\tilde{\alpha}$, 
    the two integrals are separable.
We first execute the integral over 
    $\tilde{\theta}$, which is
\begin{align}
    \frac{\rmd W_{\rm mono}}{\rmd \omega}(\omega, \theta, \gamma)
    = 2\pi
    \int_{-\theta}^{\pi-\theta}  
    \rmd \tilde{\theta} 
    \ \sin(\tilde{\theta}+\theta) \  
    \frac{\rmd W_{\rm tot}}{\rmd \omega \, \rmd \Omega}
    (\omega, \tilde{\theta}, \tilde{\theta}+\theta, \gamma) \ . 
\end{align} 
As the emission is 
    highly concentrated around $\tilde{\theta} = 0$,  
  we may consider an approximation to this integral following \cite{Westfold59ApJ} and \cite{Rybicki1979rpa}:  
\begin{align}
    \frac{\rmd W_{\rm mono}}{\rmd \omega}(\omega, \theta, \gamma)
    & \approx  2\pi 
    \int_{-\infty}^{+\infty}   
      \rmd \tilde{\theta} \ 
      \sin\theta \,
    \frac{\rmd W_{\rm tot}}{\rmd \omega \, \rmd \Omega}
    (\omega, \tilde{\theta}, \theta, \gamma) 
   \nonumber \\ & 
    =   \frac{\sqrt{3}e^3B\sin\theta}{2\pi m_e c^2}  F(x) \   , 
\end{align}  
  with $x = ({\omega}/{\omega_c})$,
    $\omega_c = (3/2)\gamma^3\omega_B \sin\tilde{\alpha}$
    and 
    $F(x) = x \int_x^{\infty}  \rmd \xi \ 
      K_{5/3}(\xi)$.
Similarly, we obtain for the linear polarisation that 
\begin{align}
    \frac{\rmd W_{{\rm mono},{\rm pol}}} 
      {\rmd \omega}(\omega, \theta, \gamma)
    & \approx  2\pi 
    \int_{-\infty}^{+\infty} \rmd \tilde{\theta} \ 
    \sin\theta \  
    \frac{\rmd W_{\rm pol}}{\rmd \omega \, \rmd \Omega}
    (\omega, \tilde{\theta}, \theta, \gamma) 
    \nonumber \\ & 
   = 
    \frac{\sqrt{3}e^3B\sin\theta}{2\pi m_e c^2} \ 
    G(x) \  ,  
\end{align} 
   where $G(x) = x K_{2/3}(x)$.   

With the approximation above, 
    the specific flux of synchrotron radiation from electrons with isotropic distribution can be simplified to
\begin{align}
    & F_\omega(\omega, \theta) = 
    \frac{1}{4\pi D^2} \frac{\sqrt{3}e^3B\sin\theta}{2\pi m_e c^2}
    \int_{\gamma_{\rm min}}^{\gamma_{\rm max}} 
    \rmd \gamma \,
    F(\omega/\omega_c) 
    \frac{\rmd N_e(\gamma)}{\rmd \gamma} \ ; 
    \label{eq:iso_tot}
    \\
    & F_{\omega,{\rm pol}} (\omega, \theta) = 
    \frac{1}{4\pi D^2} \frac{\sqrt{3}e^3B\sin\theta}{2\pi m_e c^2}
    \int_{\gamma_{\rm min}}^{\gamma_{\rm max}} 
    \rmd \gamma \,
    G(\omega/\omega_c) 
    \frac{\rmd N_e(\gamma)}{\rmd \gamma} \ ,
    \label{eq:iso_pol}
\end{align}
where $N_e(\gamma, \tilde{\alpha}) = 
    (4\pi)^{-1}(\rmd N_e/\rmd \gamma)$ for isotropic distributions.
For anisotropic distributions, 
   it is necessary to use 
    equation~\ref{eq:F_omega} and \ref{eq:F_pol_omega}. 
In astronomy, 
  the synchrotron fluxes  are more commonly expressed 
    in term of frequency $\nu$ 
    instead of the angular frequency $\omega$, 
    and  
    $F_\nu(\nu) = 2\pi F_\omega(\omega)$
    with $\nu = \omega/(2\pi)$.

%
\subsection{Electron energy distribution functions} 
\label{subsec: e-energy}  

We consider an electron distribution function
    that depends on the particle energy and the momentum pitch angle 
    with respect to the magnetic field. 
We consider a function $N_e(\gamma, \tilde{\alpha})$ that 
    can be decomposed into 
two functions  
   $f_e(\gamma)$ and $g_e(\tilde{\alpha})$, 
   i.e.    
\begin{align}
    \frac{\rmd N_e(\gamma, \tilde{\alpha})}
    {\rmd \gamma \, \rmd \Omega}
    =  N_{e, 0} \ 
    f_e(\gamma)\, g_e(\tilde{\alpha}) \ , 
\end{align} 
where $N_{e,0}$ 
  is the total number of electrons 
  in the entire emitting volume, 
and the normalisations are 
\begin{align} 
 \int_1^{\infty} \rmd \gamma\,
  f_e(\gamma)  = 
 2 \pi \int_0^\pi \rmd \tilde{\alpha}\,  
   \sin\tilde{\alpha} \ 
  g_e(\tilde{\alpha})   
    = 1 \ .  
\end{align}  
In this study we consider the generic cases of power-law,
  broken power-law and double power-law distributions 
  and two additional cases of kappa and log-parabola distributions  
  for their relevance for astrophysical applications. The 
    kappa distribution was introduced 
      to provide fits to data from 
      the solar wind observations. 
      It was later adopted by astrophysicists, 
      as it smoothly connects 
      the thermal and non-thermal particle components 
    \citep[see e.g.][]{Davelaar2018A&A,Fromm2022A&A,EHT2022ApJ}. 
The log-parabola distribution is used to provide 
    fits to the blazar 
    emission spectrum in some studies
\citep{Landau1986ApJ,Krennrich1999ApJ,Massaro2004A&A,Dermer2014ApJ}.
This distribution can be generated 
  by certain particle acceleration mechanisms 
  \citep[see][]{Kardashev1962SvA, Massaro2004A&A, Tramacere2007A&A}.  
In addition, we calculate the synchrotron spectrum assuming all electrons have the same energy. 
This case is unrealistic for astrophysical applications, but an interesting theoretical case study. 
The result is presented in Appendix~\ref{app:mono}.
  \\ 


\noindent 
\textbf{(i) Power law:} \\  
An extended power law is one of the generic particle distributions 
  used in astrophysical calculation. 
We adopt 
\begin{align}
    f_e(\gamma) =
    \frac{(p-1)}{{\gamma_1}^{1-p}-{\gamma_2}^{1-p}}
    \gamma^{-p}
\end{align} 
  for the power-law energy distribution 
  in our calculations, with the particle energies   
   $\gamma_1 \leq \gamma \leq \gamma_2$ and 
   the power-law index $p>1$. \\

\noindent
\textbf{(ii) Broken power law and double power law:} \\ 
Often in many astrophysical situations, 
  the particle energy spectrum 
  could not be represented by an 
  extended power law. 
A common extension of the extended power-law model 
  is a spectrum jointed by two power laws with different energy spectral indices.
Such spectrum can be represented mathematically as 
\begin{align}
    f_e(\gamma) = 
    \frac{1}{\gamma_b} & 
    \left( 
    \frac{{\gamma_b}^{p_1-1}-1}{p_1 - 1} + 
    \frac{1}{p_2 - 1}
    \right)^{-1} \nonumber \\
    & \times
    \left[ \,  
    {\rm H}(\gamma_b - \gamma) \left(\frac{\gamma}{\gamma_b} \right)^{-p_1}
    + 
    {\rm H}(\gamma - \gamma_b) 
    \left( 
    \frac{\gamma}{\gamma_b}
    \right)^{-p_2}
    \right] 
\end{align}
(for $\gamma \geq 1$ and $p_1, p_2 > 1$), where   
   ${\rm H}(x)$ is the Heaviside step function, 
    whose value is 0 when $x < 0$ and 1 when $x \geq 0$. 
For energy spectral index $p$ taking  
  a larger value at higher energies 
  ($p_2 > p_1$), 
  the spectrum is generally referred to as a broken power law. 
A broken power law could be intrinsic to the acceleration process \citep[see e.g.,][]{Comisso2024ApJ}, or develop gradually from an initially single power law 
due to more efficient cooling of the higher-energy particles, e.g., synchrotron cooling.
A particle index $p$ 
   taking a smaller value at higher energies
  ($p_2 < p_1$), on the other hand, 
   may indicate 
   the presence of two populations of particles both
   with a power-law energy distribution
   but with a different energy spectral index. \\ 

\noindent
\textbf{(iii) Kappa distribution:} \\  
We may express $f_e$ in the kappa distribution as  
\begin{align}
    f_e(\gamma) =
    \Lambda
    \gamma \sqrt{\gamma^2 - 1} 
    \left(
    1 + \frac{\gamma - 1}{\kappa w}
    \right)^{-(\kappa + 1)} \ 
\end{align} 
\citep[see e.g.][]{Pandya2016ApJ},
where $\Lambda$ is a constant such that 
  $\int_1^{\infty} \rmd \gamma\,
  f_e(\gamma) = 1 $.
For a sufficiently large $\gamma$, 
  it resembles a power-law distribution with $p = \kappa - 1$.
For small $\gamma$, 
  it resembles a thermal-like distribution with its
  width regulated by the parameter $w$.  \\

\noindent
\textbf{(iv) Log-parabola distribution:} \\ 
We may express the log-parabola distribution as 
\begin{align}
    f_e(\gamma) = 
    \sqrt{\frac{2}{\pi\sigma_e^2}} 
    \left[ 
    \frac{1}{{\rm erf}(y_2) - {\rm erf}(y_1)} \right] 
    \frac{1}{\gamma} \ 
    \exp{
    \left( 
    \frac{-(\ln(\gamma)-\ln(\gamma_c))^2}{2{\sigma_e}^2}
    \right)} \ , 
\end{align}
  where $\gamma_1 \leq \gamma \leq \gamma_2$.
The parameter  
  $y_i = (\ln(\gamma_i)-\ln(\gamma_c))/\sqrt{2{\sigma_e}^2}$, with 
$i = $ 1 or 2.

\begin{figure}
    \centering
    \includegraphics[width=0.8\linewidth]{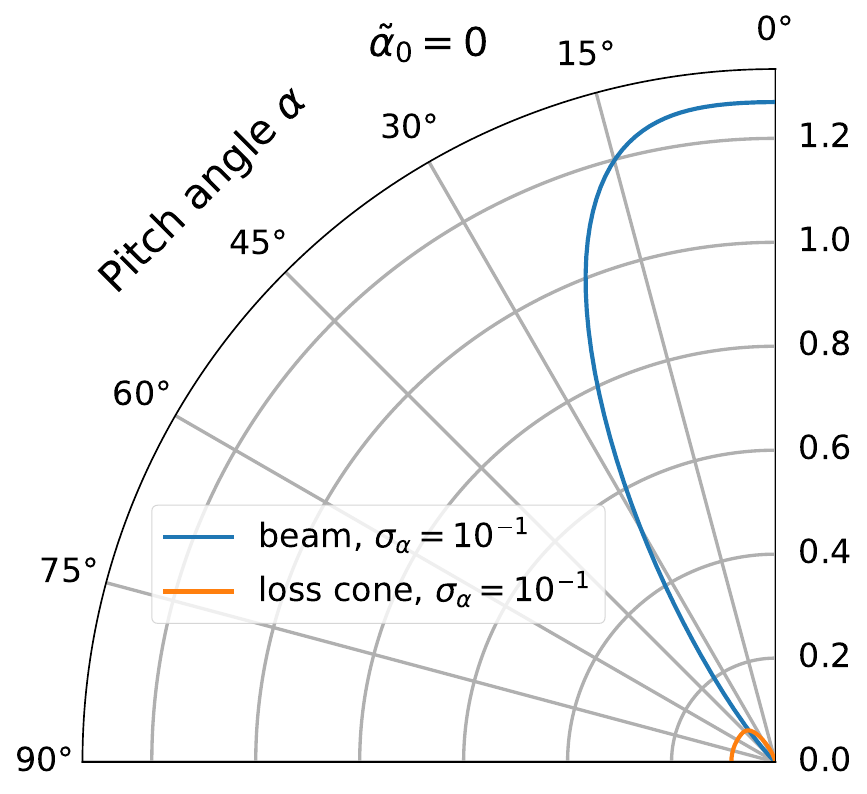}
    \includegraphics[width=0.8\linewidth]{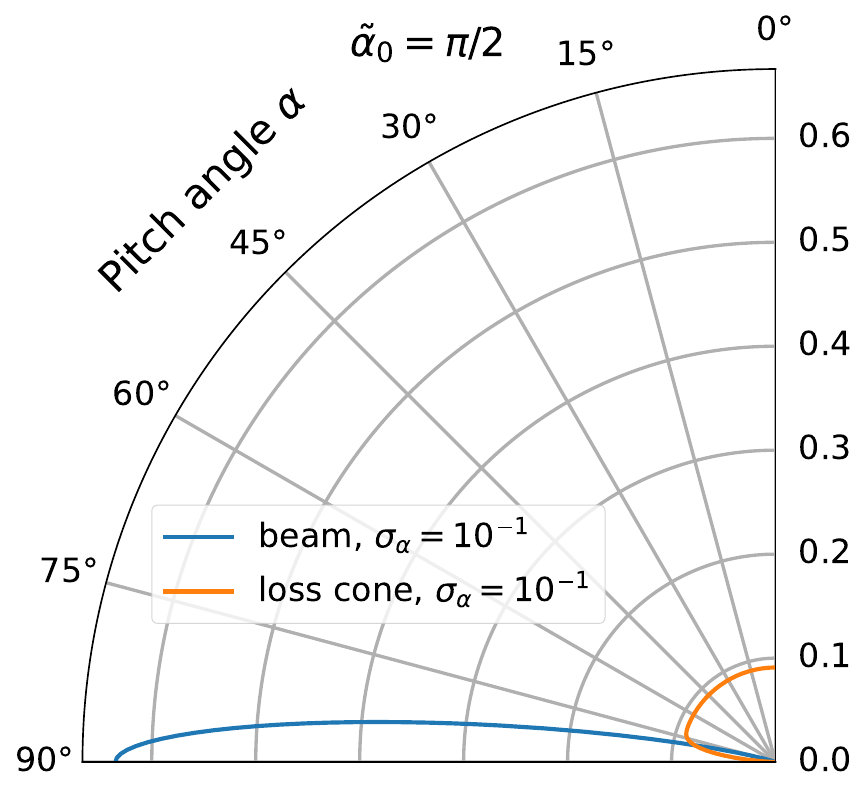}
    \caption{The pitch-angle distribution models
      used in this paper. 
    $\tilde{\alpha}_0$ specifies 
      the direction 
      of the particle beam or the loss cone; 
      $\sigma_\alpha$ determines the
    width of the beam or the loss cone.
    The upper panel shows a beam and a loss cone pointing along the polar axis ($\tilde{\alpha}_0 = 0$).
    The lower panel shows a beam and a loss cone, 
    with its axis pointing towards the equatorial plane ($\tilde{\alpha}_0 = \pi/2$).}
\label{fig:pitch_angle}
\end{figure}

%
\subsection{Electron momentum pitch-angle distribution functions} 
\label{subsec: pitch_angle}   

For the pitch-angle distributions, 
  we consider the idealised case of isotropic distribution 
  and two parametric cases of anisotropic distributions, 
  the beamed particles and the loss cone. 
Isotropic distribution is often the default assumption 
   in calculations, 
   when we have no knowledge about the  
   pitch-angle distribution.  
It is justified also in the situation 
  that particle collisions are sufficiently efficient 
  to suppress the development of pitch-angle anisotropy.   
When the particles are preferentially  
  accelerated towards a particular direction,
  it results in a particle beam.
This can occur, for example, 
  in magnetic reconnection,
    particles get stronger acceleration along
    the magnetic field direction, causing
    a beam-like distribution 
    \citep[see][]{Comisso2019ApJ, Comisso2023ApJ}. 
A loss-cone distribution could be developed  
  in magnetic mirrors or in converging magnetic-field configurations 
   \citep[][]{Baldwin1977RvMP} 
     where particles with large momentum 
    pitch angles will be trapped 
    but particles with momenta 
    roughly aligned with the magnetic field orientation  
    could escape from the confinement.  
    \\ 

\noindent
\textbf{Isotropic momentum pitch-angle distribution:} \\ 
For isotropic pitch-angle distribution, 
   the function $g_e(\tilde{\alpha})$ is a constant: 
\begin{align}
    g_e(\tilde{\alpha}) = \frac{1}{4\pi} \ .
\end{align}   

\noindent
\textbf{Beamed particles:} \\  
For the beamed distribution, 
  it takes the following form 
\begin{align}
    g_e(\tilde{\alpha}) =
    \frac{1}{\sqrt{2\pi^3{\sigma_\alpha}^2}}
    \left[ 
    \frac{1}
    {{\rm erf}(t_2)-{\rm erf}(t_1)}\right] 
    \exp\left( 
    -\frac{(\cos\tilde{\alpha} - 
    \cos\tilde{\alpha}_0)^2}{2{\sigma_\alpha}^2} 
    \right) \ ,
\end{align}
where $t_2 = (1-\cos\tilde{\alpha}_0)/\sqrt{2{\sigma_\alpha}^2}$ and 
$t_1 = (-1-\cos\tilde{\alpha}_0)/\sqrt{2{\sigma_\alpha}^2}$. 
The beam width is mainly determined by $\sigma_\alpha$ but also 
    depends on the beaming angle $\tilde{\alpha}_0$.
We list the half width at half maximum (HWHM) of the beam
    $\alpha_{\rm HWHM}$ for the parameter values 
    used in our calculations 
    (see Table~\ref{tab:HWHM}). 
For $\tilde{\alpha}_0 = 0$,
    we have 
    $\alpha_{\rm HWHM} = 0.15, 0.27, 0.49$ for 
    $\sigma_\alpha = 10^{-2}, 10^{-1.5}, 10^{-1}$,
    respectively.
For $\tilde{\alpha}_0 = \pi/2$,
    we have
    $\alpha_{\rm HWHM} = 0.012, 0.037, 0.12$ for 
    $\sigma_\alpha = 10^{-2}, 10^{-1.5}, 10^{-1}$,
    respectively.
We note that the mathematical expression of this beamed distribution is not direct results from kinetic modelling 
  but we adopt it for illustration purpose.   
It is similar to that used in \cite{Yang2018ApJ}, 
  thus allowing us to make comparison 
   with their work.
We also calculated the synchrotron spectrum using the same beamed model in \cite{Yang2018ApJ}, and the result is presented in Appendix~\ref{app:yang_zhang}.
    
\begin{table}
    \centering
    \begin{tabular}{|c|c|c|c|}
        \hline
         & $\sigma_\alpha = 10^{-2}$ & $\sigma_\alpha = 10^{-1.5}$ & $\sigma_\alpha = 10^{-1}$ \\
        \hline
        $\tilde{\alpha}_0 = 0$ & $\alpha_{\rm HWHM} = 0.15$ & 0.27 & 0.49 \\
        \hline
        $\tilde{\alpha}_0 = \pi/2$ & 0.012 & 0.037 & 0.12 
        \\
        \hline
    \end{tabular}
    \caption{The HWHM of the beam, $\alpha_{\rm HWHM}$, 
    for different values of $\tilde{\alpha}_0$ and $\sigma_\alpha$ 
    adopted in the calculations.}
    \label{tab:HWHM}
\end{table}    

\noindent
\textbf{Loss cone:} \\ 
The loss-cone distribution may be considered 
  as a complement of a beamed distribution.  
We modify the beamed distribution above
    and model the loss-cone distribution as 
\begin{align}
    g_e(\tilde{\alpha}) = &
    \left(
    4\pi - \sqrt{2\pi^3{\sigma_\alpha}^2}\ 
    \big[{\rm erf}(t_2) - {\rm erf}(t_1)\big]
    \right)^{-1} \nonumber \\
    & \times
    \left(
    1 - \exp\left( 
    -\frac{(\cos\tilde{\alpha} - \cos\tilde{\alpha}_0)^2}{2{\sigma_\alpha}^2} 
    \right)
    \right) \ .
\end{align}  \\ 
The size of the loss cone is the same as that in
    the beam model, which is shown in 
    Table~\ref{tab:HWHM}.
We show in Figure~\ref{fig:pitch_angle} the polar plots  
   of examples of beamed and loss-cone pitch-angle distributions. 

\begin{figure*}
    \vspace*{0.5cm}
    \centering
    \includegraphics[width=0.33\linewidth]{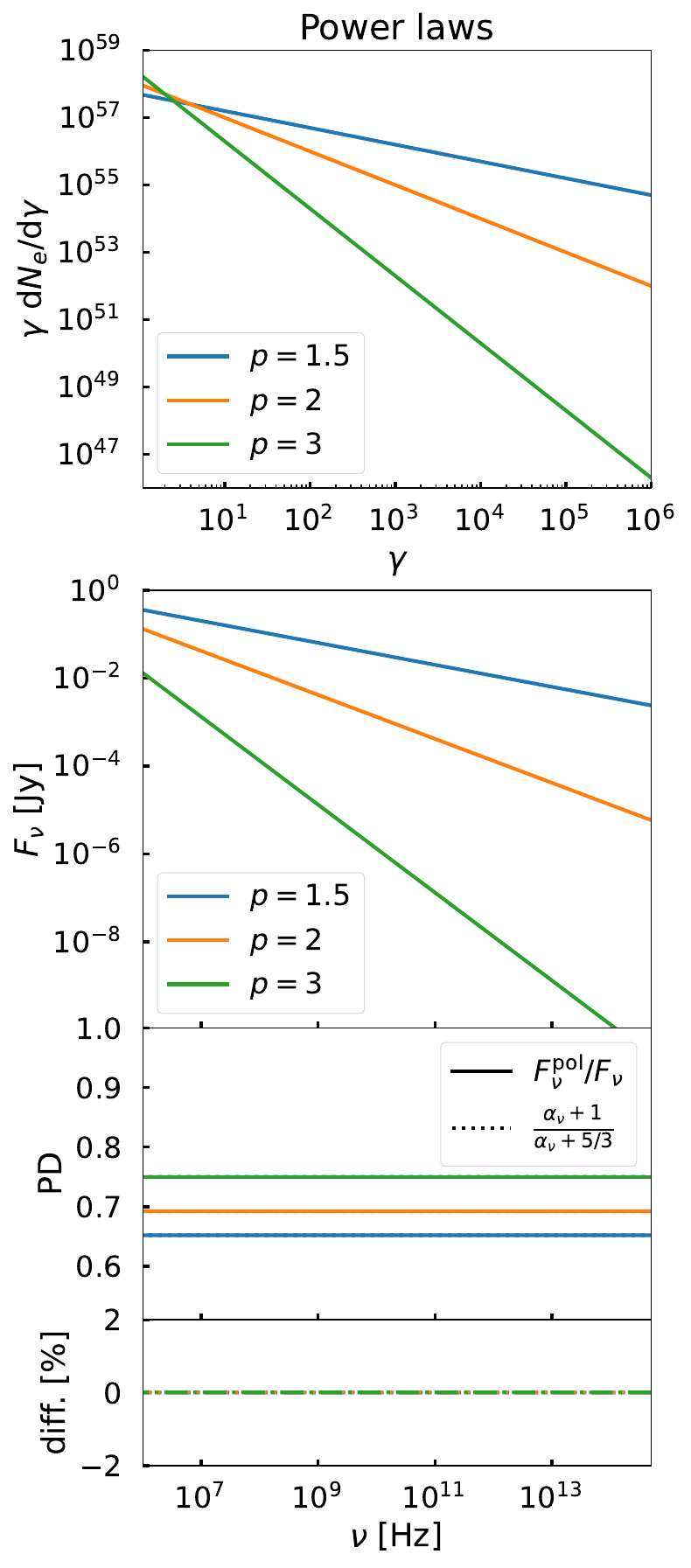}
    \includegraphics[width=0.33\linewidth]{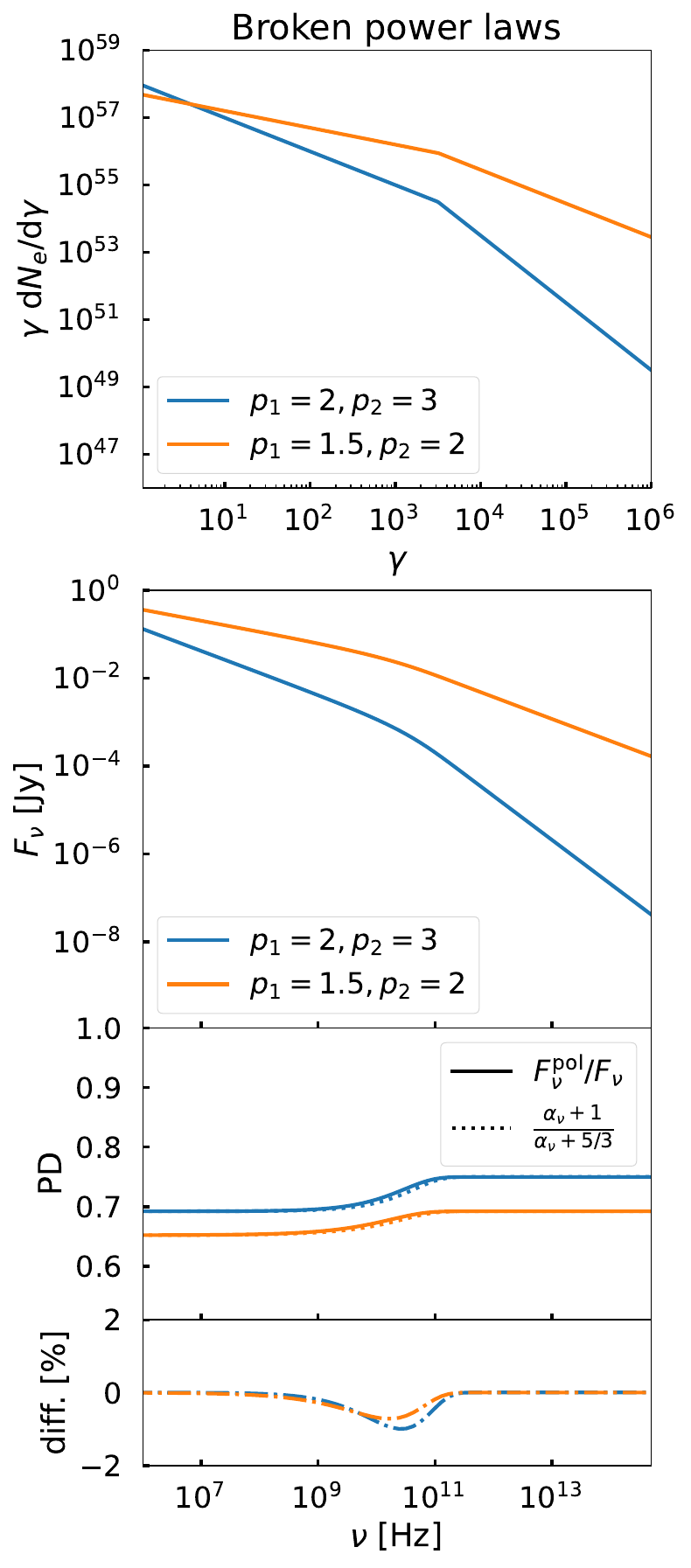}
    \includegraphics[width=0.33\linewidth]{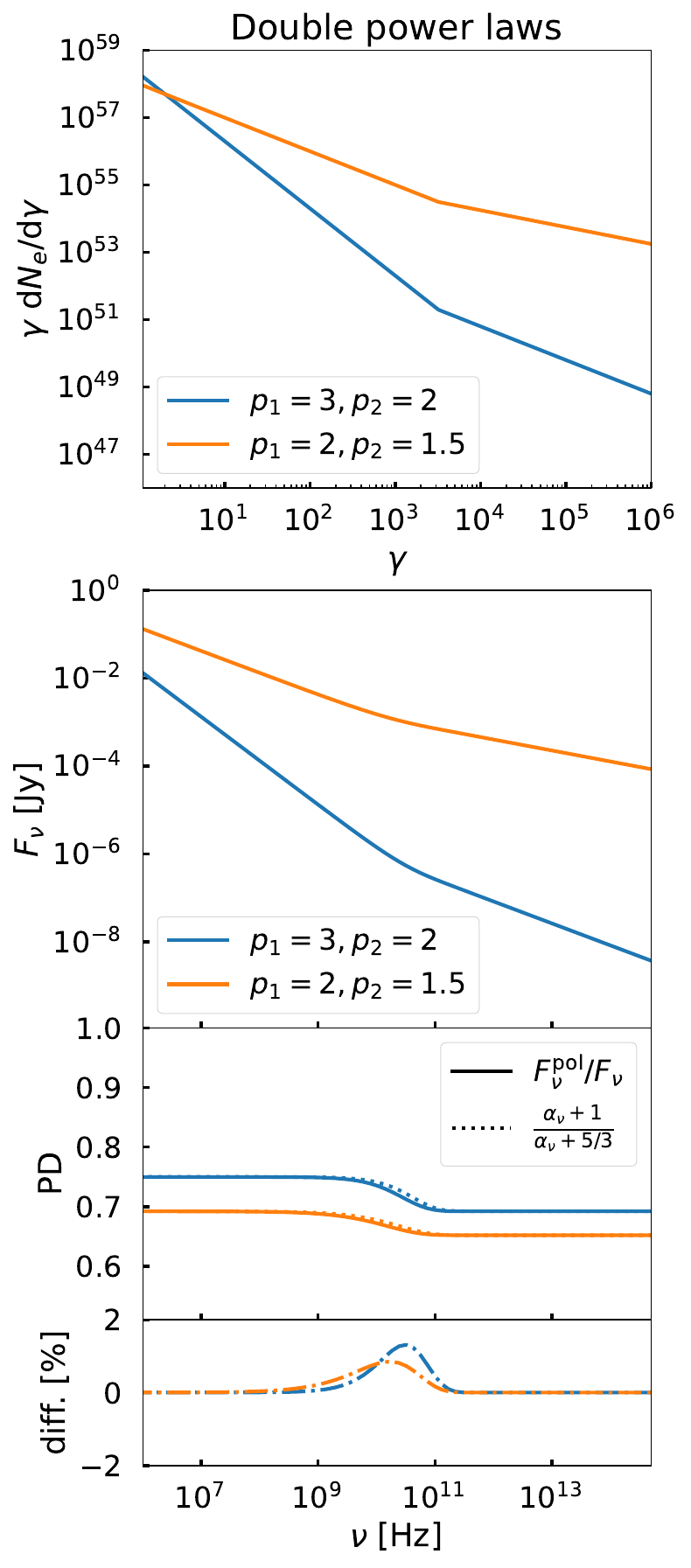}
    \caption{The three columns show the synchrotron spectra and polarisation properties of three particle energy distributions, which from left to right are single power-law, broken power-law, and double power-law, assuming an isotropic pitch-angle distribution. 
    The first row shows the particle energy distributions.
    All the distributions extends from $\gamma = 1$
    to infinity.
    The second row shows the corresponding synchrotron spectrum.
    The third row shows the frequency-dependent PD.
    PD is calculated using two different methods.
    The solid line is calculated using the full formula (Eq.~\ref{eq:iso_tot} and \ref{eq:iso_pol}), and the dotted line is calculated from
    the local spectral index $\alpha_\nu$ (Eq.~\ref{eq:pd_gen}).
    The fourth row shows the percentage difference
    of the PD results between these two methods.
    The following parameters are used:
    $\gamma_b = 10^{3.5}$, 
    $\theta = \pi/2$,
    $N_{e, 0} = 10^{58}$, $B = 1\,$mG, and $D = 1$\,Gpc.}
\label{fig:pl_inf}
\end{figure*}

\begin{figure*} 
\vspace*{0.5cm}
    \centering
    \includegraphics[width=0.33\linewidth]{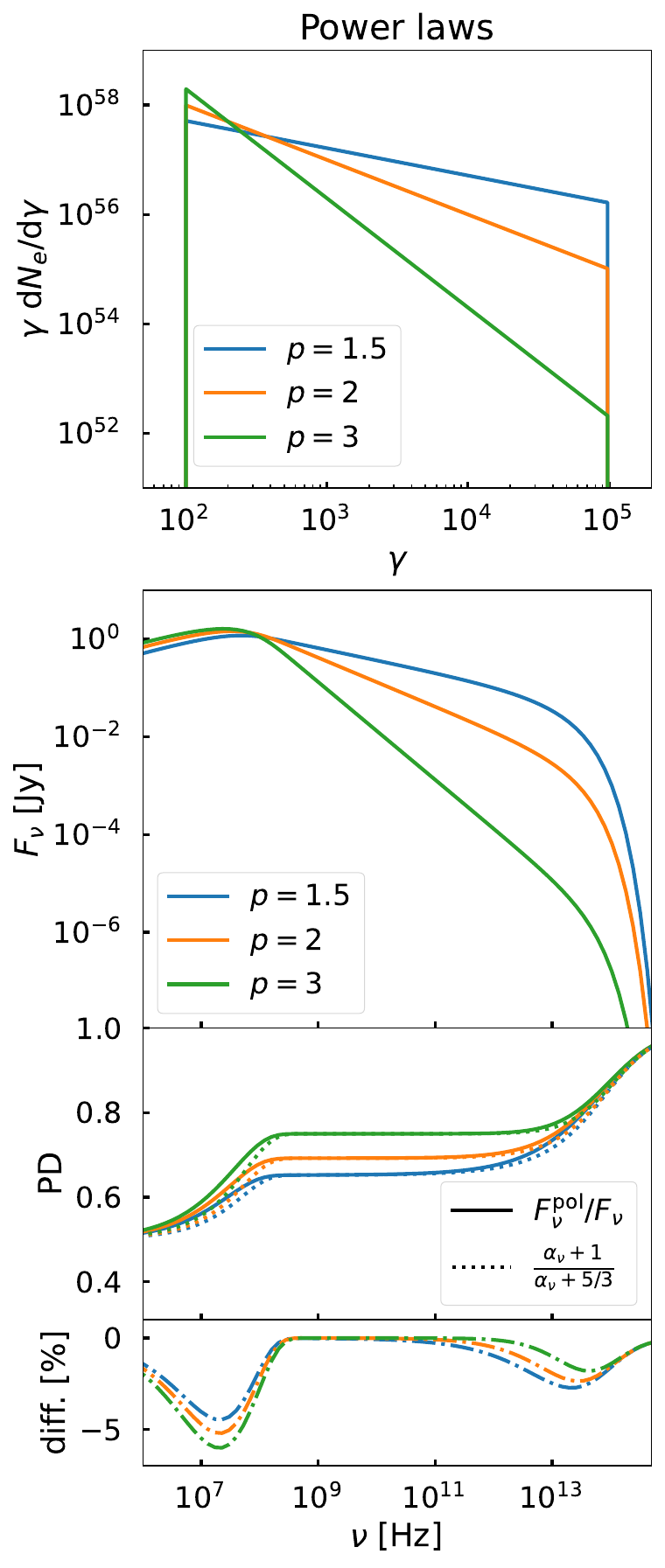}
    \includegraphics[width=0.33\linewidth]{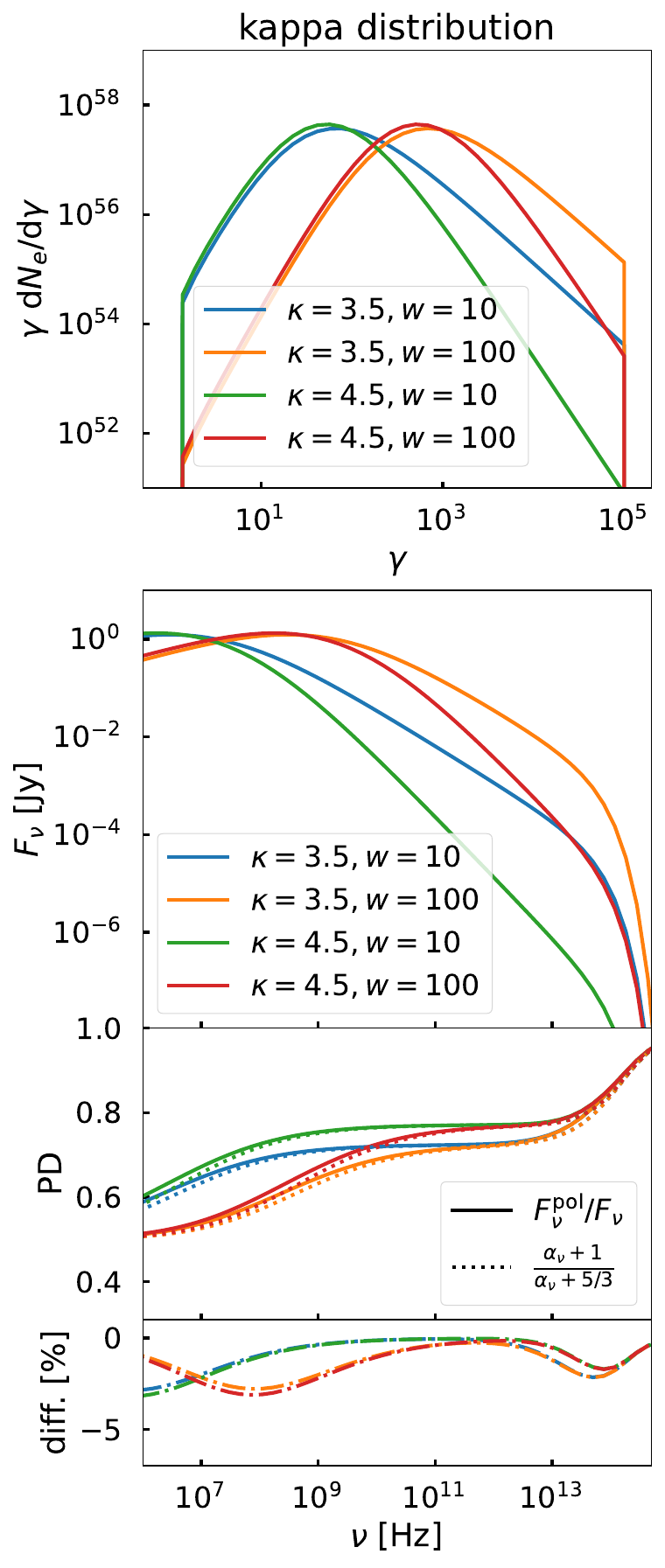}
    \includegraphics[width=0.33\linewidth]{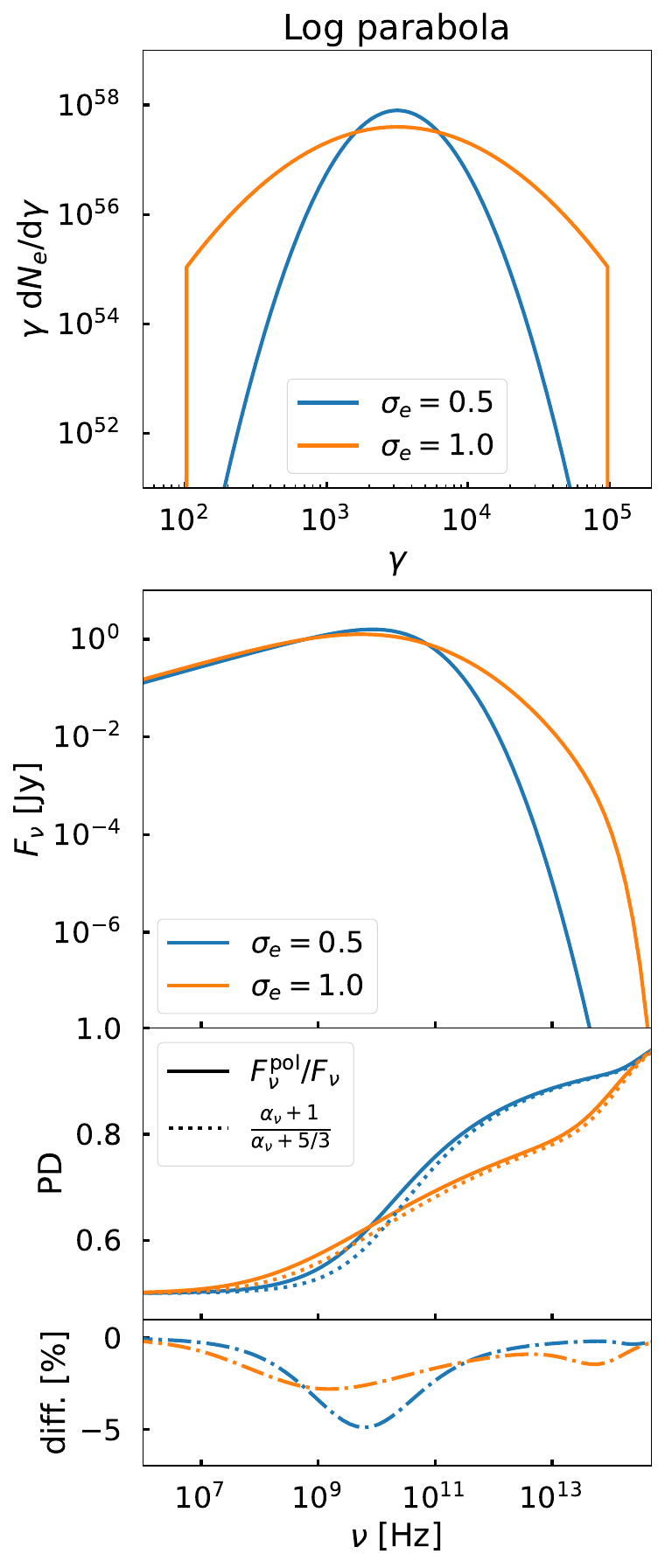}
    \caption{The three columns show the synchrotron spectra and polarisation properties of three particle energy distributions, which from left to right are single power-law, kappa, and log-parabola, assuming an isotropic pitch-angle distribution. 
    The first row shows the particle energy distributions.
    All the distributions have a lower
    cut-off $\gamma_1 = 10^2$ and higher energy cut-off $\gamma_2 = 10^5$.
    The second row shows the corresponding synchrotron spectrum.
    The third row shows the frequency-dependent PD.
    PD is calculated using two different methods.
    The solid line is calculated using the full formula (Eq.~\ref{eq:iso_tot} and \ref{eq:iso_pol}), and the dotted line is calculated from
    the local spectral index $\alpha_\nu$ (Eq.~\ref{eq:pd_gen}).
    The fourth row shows the percentage difference
    of the PD results between these two methods.
    The following parameters are used:
    $\gamma_c = 10^{3.5}$,
    $\theta = \pi/2$, 
    $N_{e, 0} = 10^{58}$, $B = 1\,$mG, and $D = 1$\,Gpc.}
\label{fig:cutoff}
\end{figure*}

%
\section{Results}
\label{sec:results} 

Without losing generality,  
 we adopt the followings as fiducial values of 
 the total number of electrons, the magnetic field strength, and the distance between the emitting object and the observer in our calculations:  
    $N_{e, 0} = 10^{58}$, $B = 1\,{\rm mG}$, 
    and $D = 1\, {\rm Gpc}$. 
While these are specific values in the calculations, 
  extensive variables such as flux (and flux density) 
  can be obtained by direct scaling. 
On the other hand, PD is an intensive quantity,
    which is independent of $N_{e, 0}$, $D$, and 
    also the observing frequency $\nu$ when normalised with respect to $\nu_c$.

%
\subsection{Isotropic pitch-angle distributions}  \label{sec:isotropic}

Figure~\ref{fig:pl_inf} shows  
 the particle distributions, 
 the emission spectra, in terms of the specific flux $F_\nu$,  
 and the degree of linear polarisation,  
 ${\rm PD} (\equiv \Pi_{\rm L})$ 
 (in panels from top to bottom) 
 of synchrotron radiation from relativistic electrons 
 with an isotropic pitch-angle distribution. 
We found that the degree of linear polarisation is
    well described by
    a generalised version of Eq.~\ref{eq:pd_pl},
\begin{align} \label{eq:pd_gen}
    \Pi_{\rm L,gen} = \frac{\alpha_\nu+1}{\alpha_\nu+5/3}
\end{align}
    where $\alpha_\nu = -(\rmd \log{F_\nu}/\rmd \log{\nu})$ 
    is the local spectral index,
  for broken power-law and double power-law 
  energy distributions  
  as in the canonical case of 
  the extended power-law distribution. 
We refer to Eq.~\ref{eq:pd_gen} as the generalised PD formula or the generalised formula hereafter.
There is only a small discrepancy, 
  of less than 2\%, 
  at the transition between the two power laws 
  in the broken power-law and double power-law 
  energy distributions. 
Figure~\ref{fig:cutoff} 
  shows the correspondences 
  for three energy distributions: 
  a power law with a high-energy and a low-energy cut-off, 
  kappa distribution and log-parabola distributions. 
As in the cases of extended power-law, broken power-law 
 and double power-law energy distribution, 
 the same expressions for $\Pi_{\rm L}$ 
 are generally applicable when the local value 
 for $\alpha_\nu$ is specified.   
The discrepancies in all cases are 
  at most $\sim$5\%.

\begin{figure*} 
    \centering
    \includegraphics[width=0.33\linewidth]{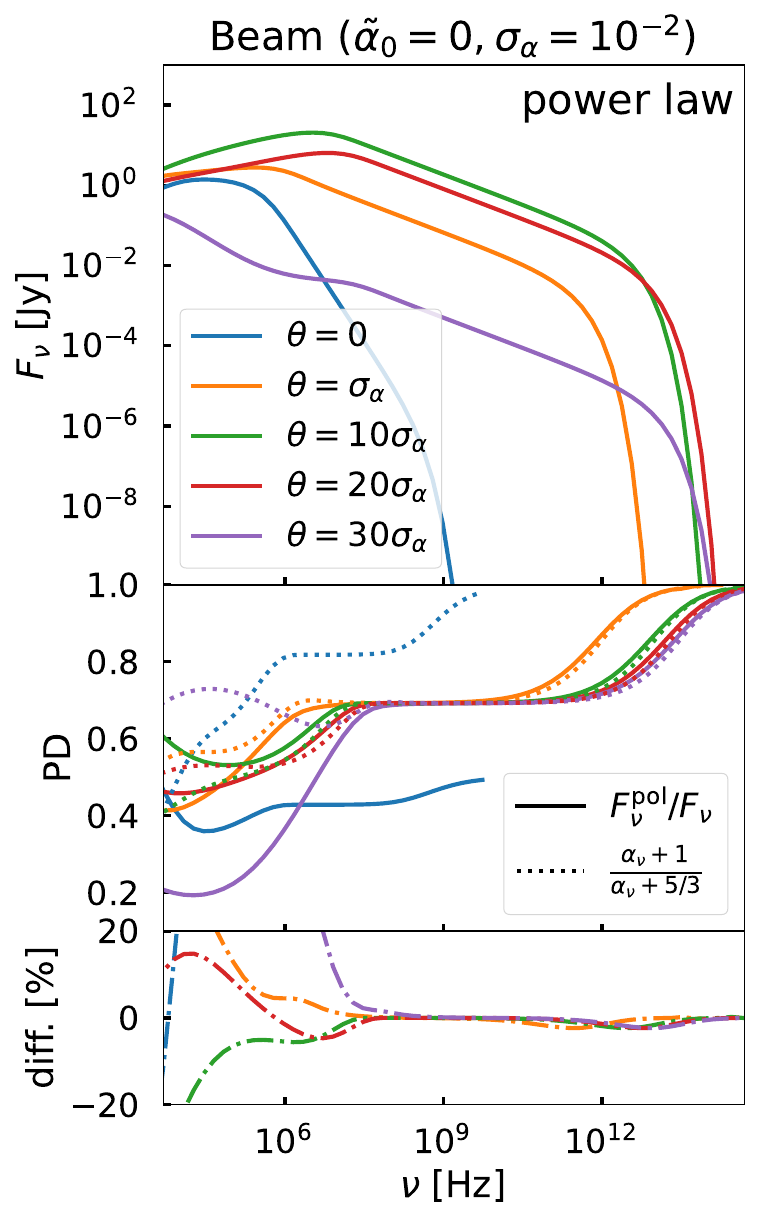}
    \includegraphics[width=0.33\linewidth]{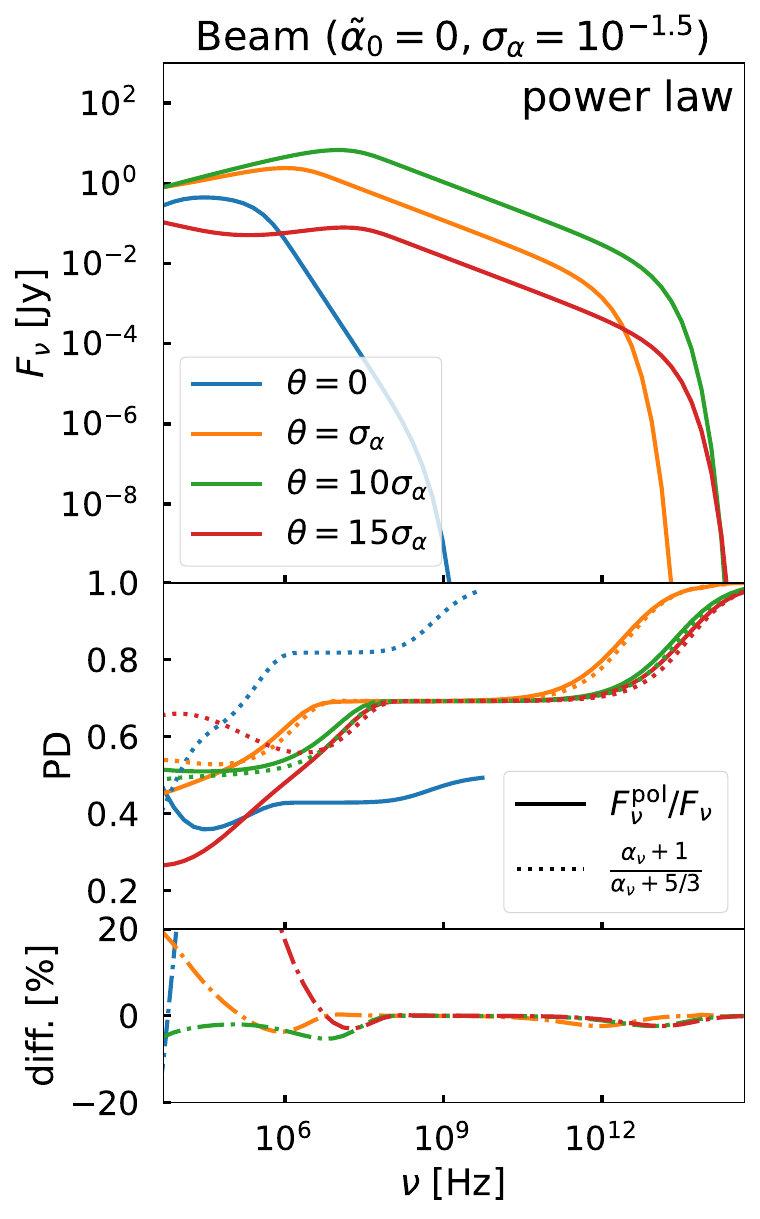}
    \includegraphics[width=0.33\linewidth]{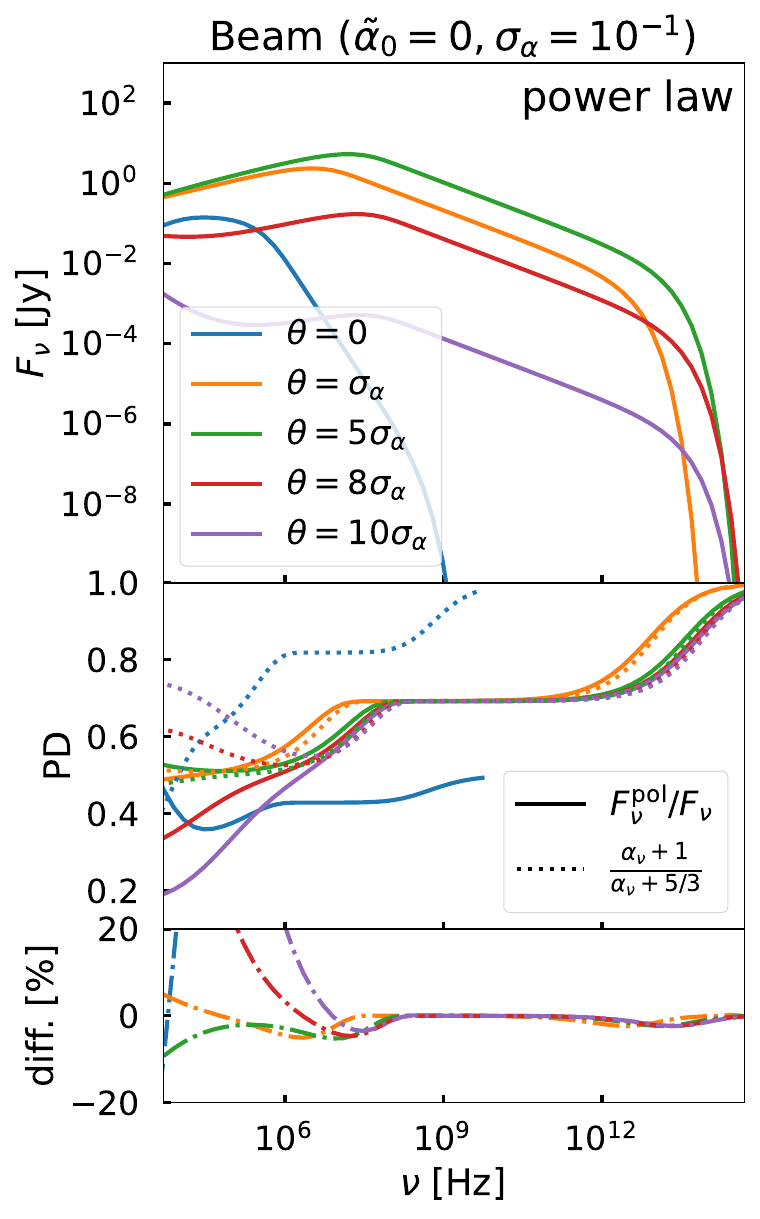}    
    \includegraphics[width=0.33\linewidth]{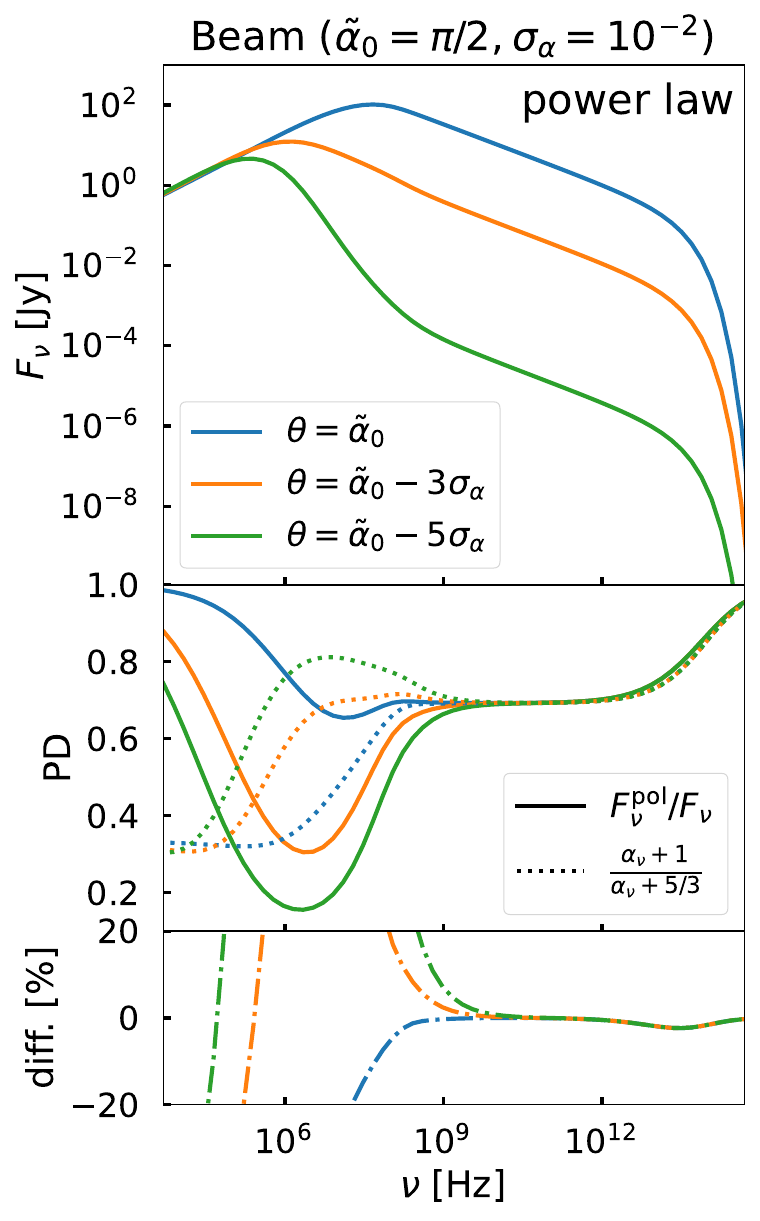}
    \includegraphics[width=0.33\linewidth]{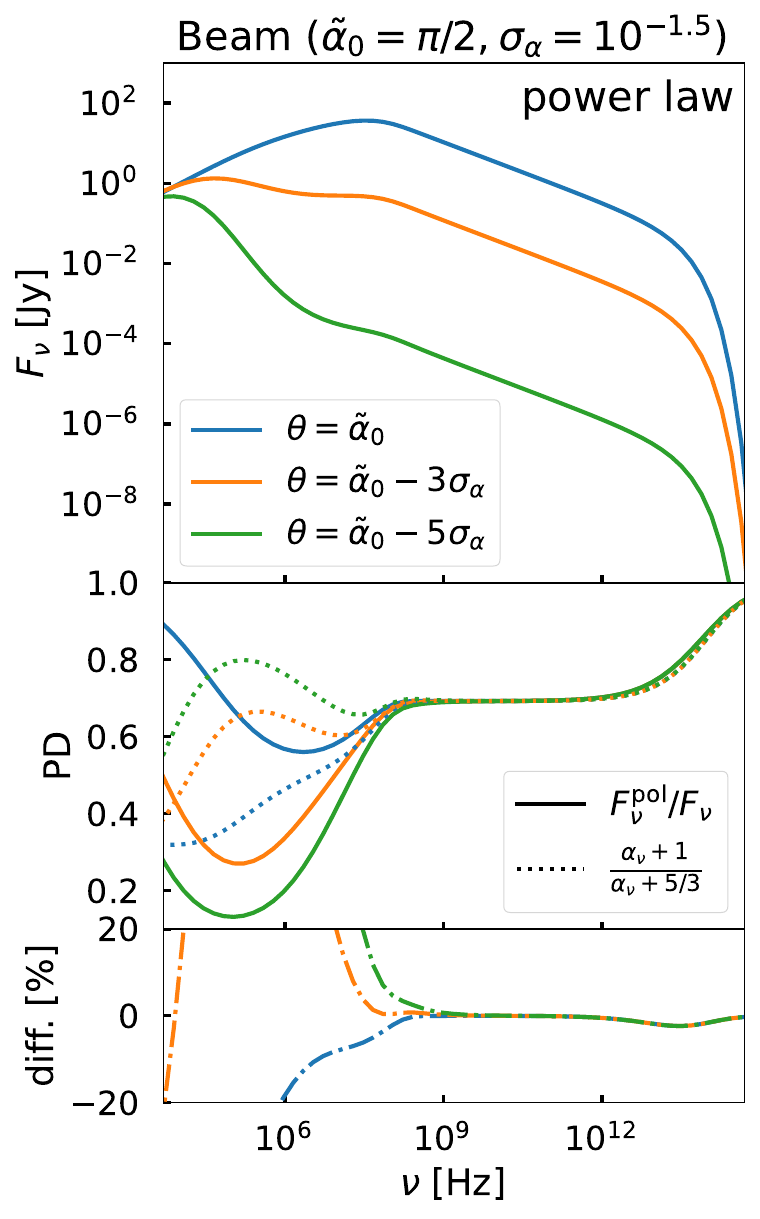}
    \includegraphics[width=0.33\linewidth]{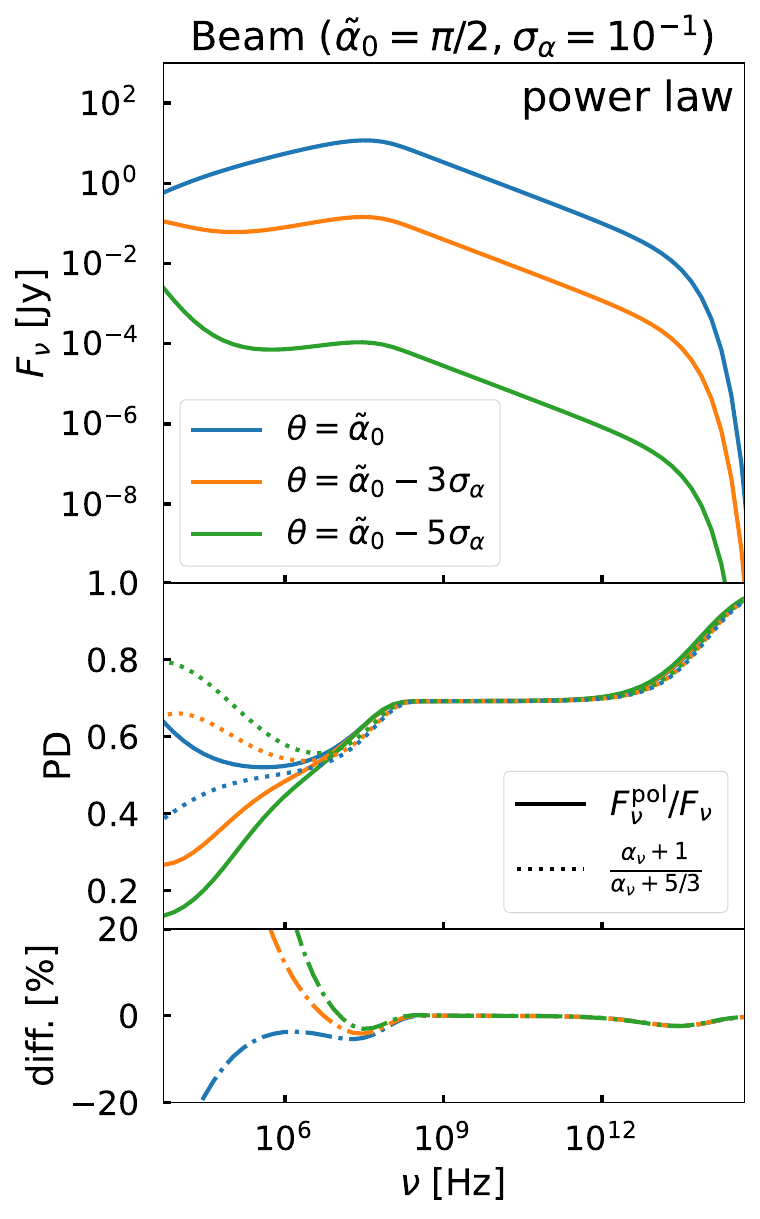}
    \caption{Synchrotron spectra of the beamed model with different values of the parameters, i.e.
    the beaming direction $\tilde{\alpha}_0$, 
    the opening angle of the beam $\sigma_\alpha$,
    and the viewing angle $\theta$.
    For all the calculations, electrons are assumed to follow
    a power-law energy distribution with index $p = 2$, a low-energy cut-off $\gamma_1=10^2$ and high-energy cut-off $\gamma_2=10^5$.
    PD is calculated using two methods.
    The solid line is calculated using the full formula (Eq.~\ref{eq:F_omega} and \ref{eq:F_pol_omega}), and the dotted line is calculated from
    the local spectral index $\alpha_\nu$ (Eq.~\ref{eq:pd_gen}).
    The beam width of these models are 
    $\alpha_{\rm HWHM} = 0.15, 0.27, 0.49, 0.012, 0.037, 0.12$ from left to right, from top to bottom.
    Besides, the following parameters are used:
    $N_{e, 0} = 10^{58}$, $B = 1\,$mG, and $D = 1$\,Gpc.}
\label{fig:beam_pl}
\end{figure*}

\begin{figure*} 
    \centering
    \includegraphics[width=0.33\linewidth]{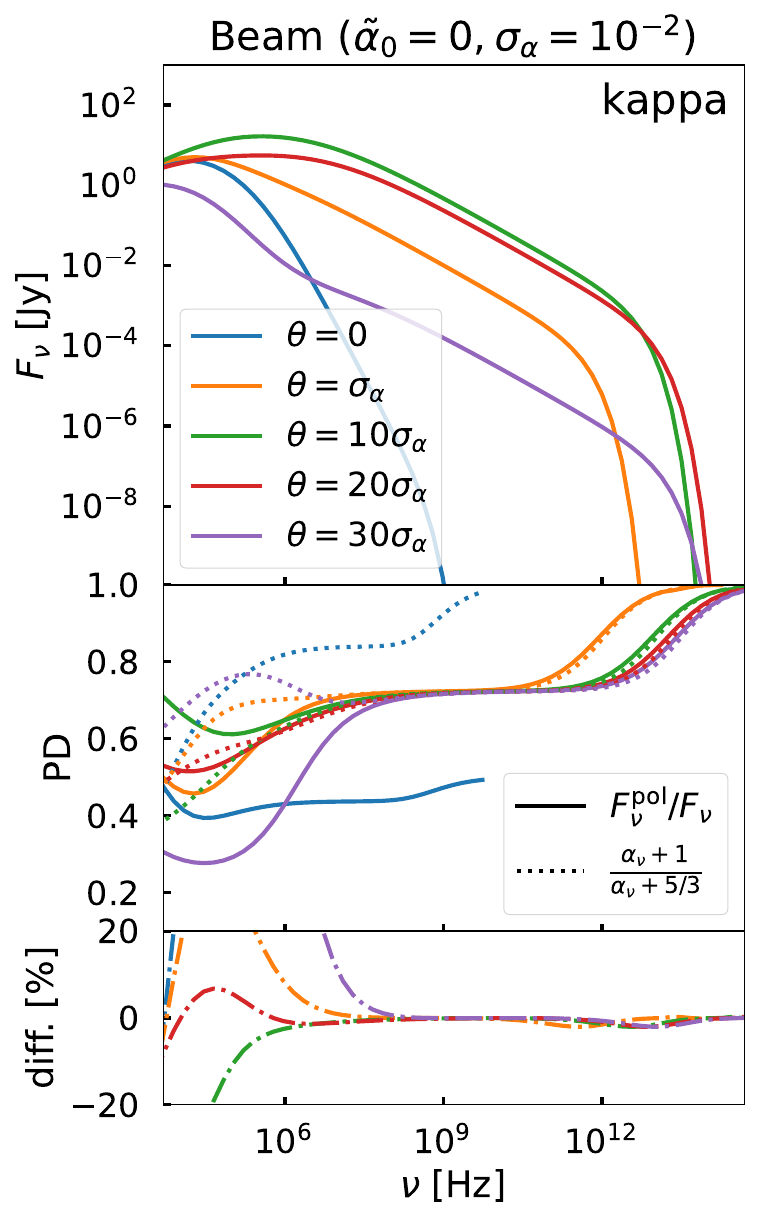}
    \includegraphics[width=0.33\linewidth]{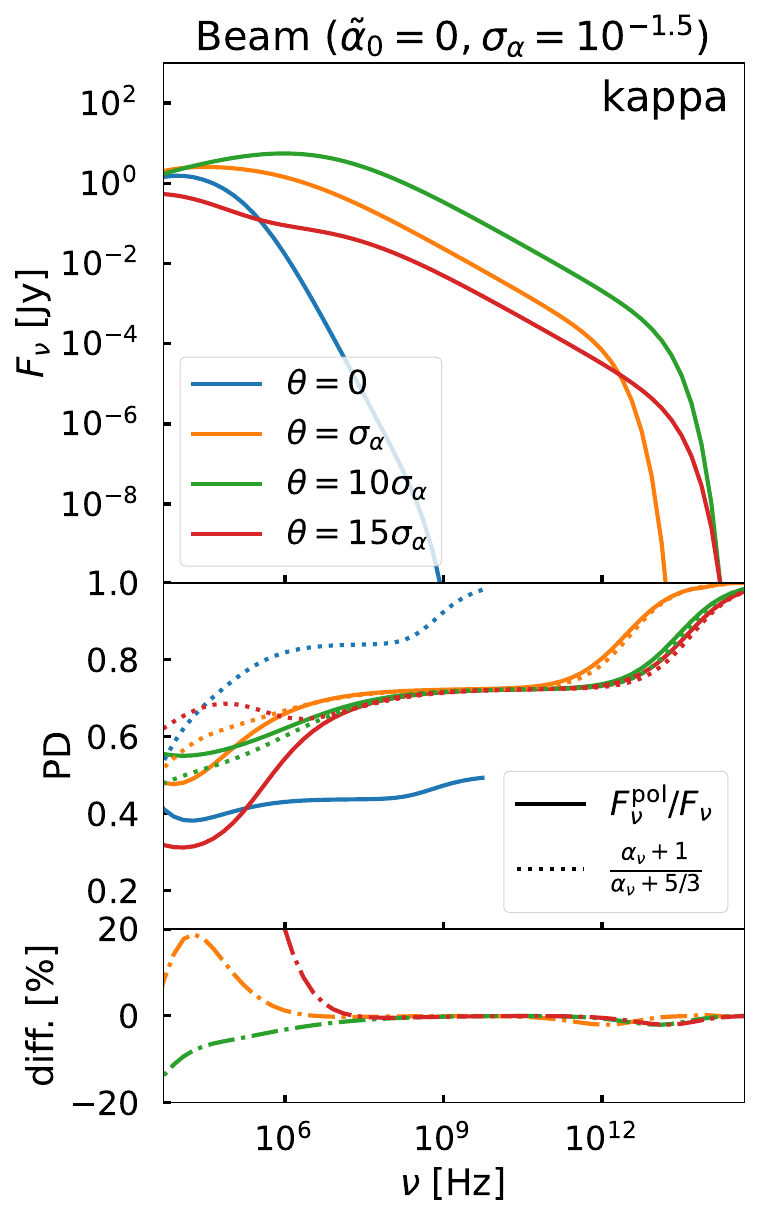}
    \includegraphics[width=0.33\linewidth]{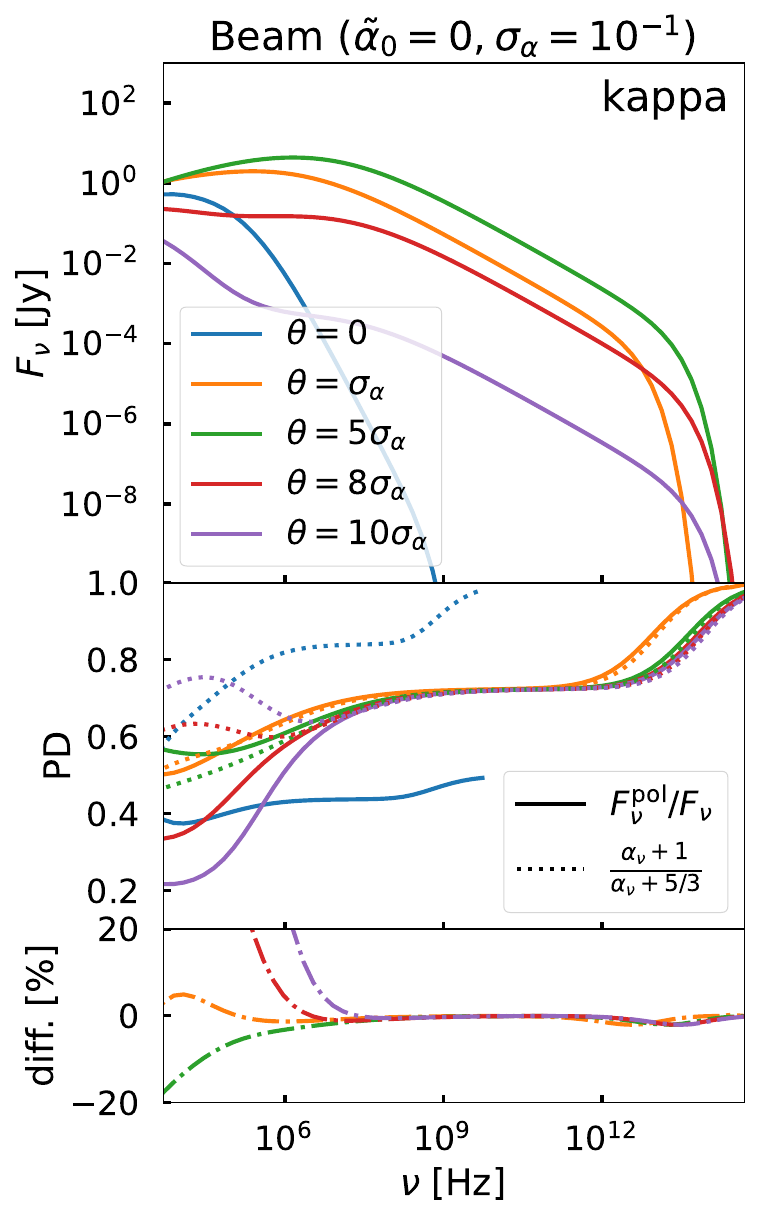}    
    \includegraphics[width=0.33\linewidth]{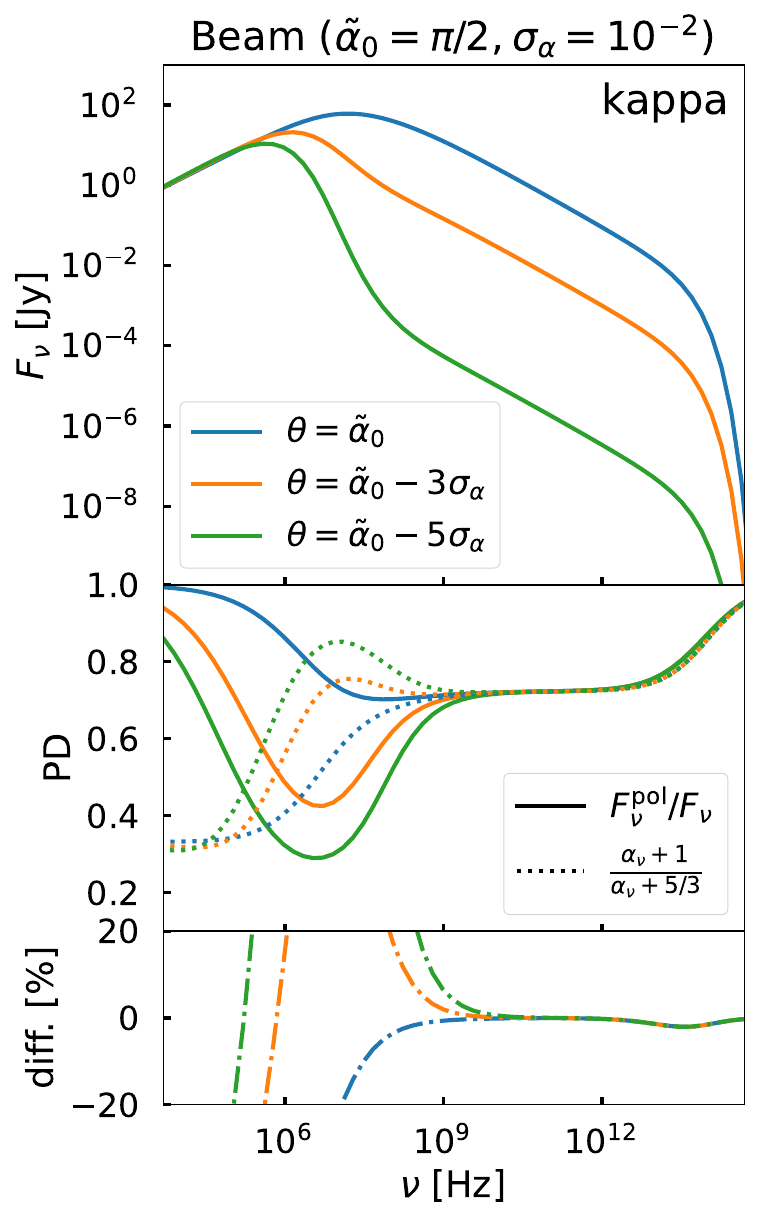}
    \includegraphics[width=0.33\linewidth]{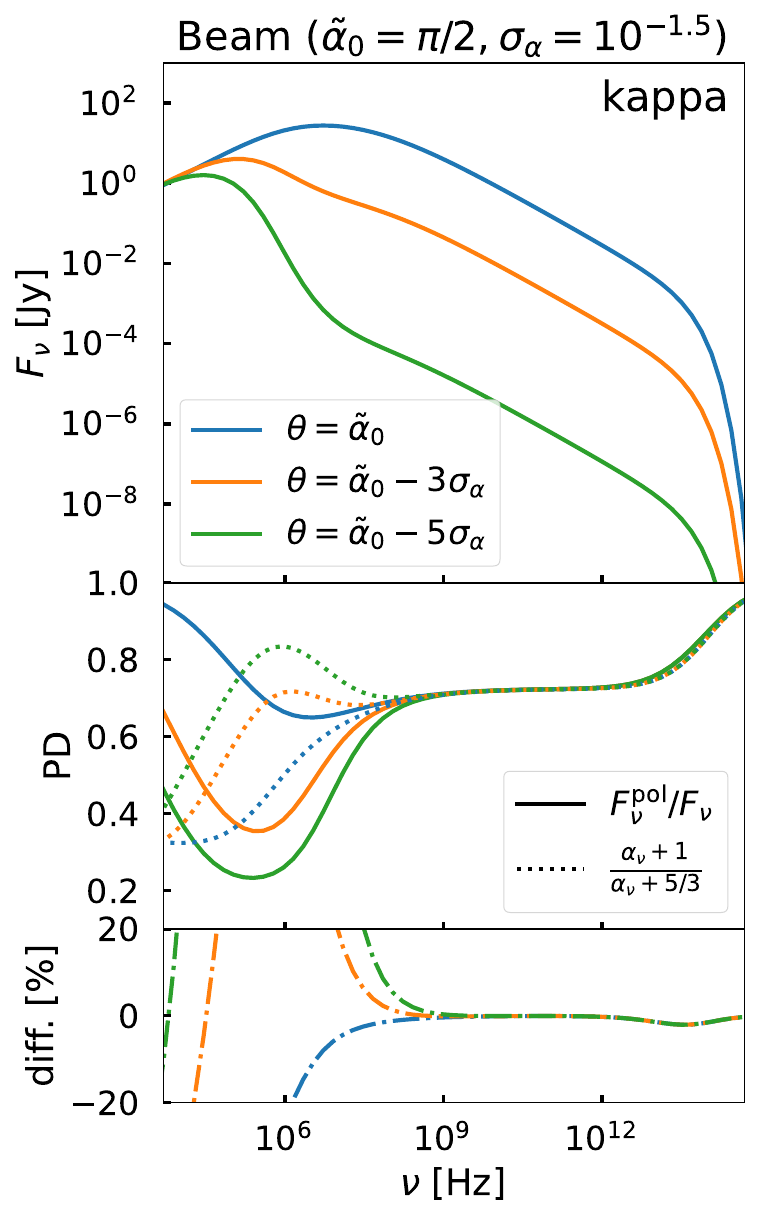}
    \includegraphics[width=0.33\linewidth]{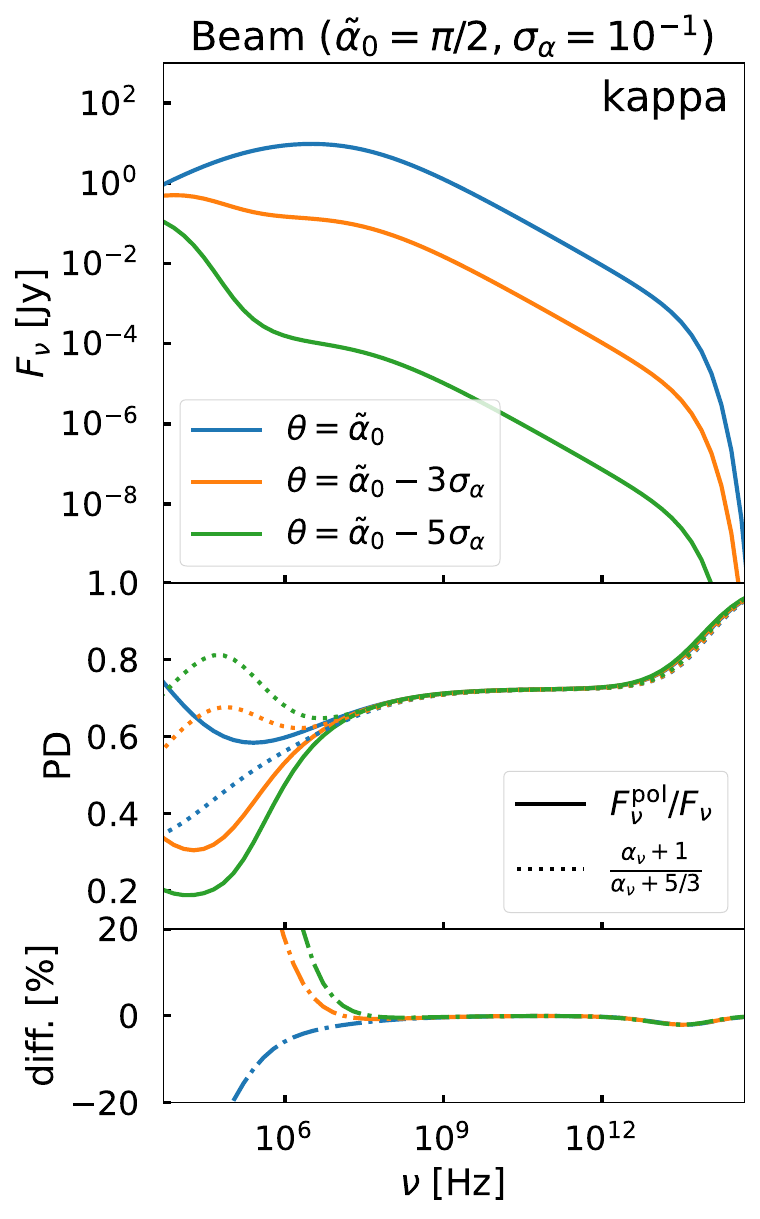}
    \caption{Synchrotron spectra of the beam model with different values of the parameters, i.e.
    the beaming direction $\tilde{\alpha}_0$, 
    the opening angle of the beam $\sigma_\alpha$,
    and the viewing angle $\theta$.
    For all the calculations, electrons are assumed to follow
    a kappa distribution with $\kappa = 3.5$ 
    and $w = 10$, and a high-energy cut-off $\gamma=10^5$.
    PD is calculated using two methods.
    The solid line is calculated using the full formula (Eq.~\ref{eq:F_omega} and \ref{eq:F_pol_omega}), and the dotted line is calculated from
    the local spectral index $\alpha_\nu$ (Eq.~\ref{eq:pd_gen}).
    The beam width of these models are 
    $\alpha_{\rm HWHM} = 0.15, 0.27, 0.49, 0.012, 0.037, 0.12$ from left to right, from top to bottom.
    Besides, the following parameters are used:
    $N_{e, 0} = 10^{58}$, $B = 1\,$mG, and $D = 1$\,Gpc.}
\label{fig:beam_kappa}
\end{figure*}

%
\subsection{Anisotropic distributions}   \label{sec:anisotropic}

Here our focus is on how the anisotropic 
 pitch-angle 
  distribution would alter the spectropolarimetry 
  properties of synchrotron radiation. 
For illustration  
  we consider only the power-law 
    and kappa distributions for the electron energies. 
In each case,  
  we calculate $F_\nu(\nu)$ and $\Pi_{\rm L}$
  for the beamed anisotropy and the loss-cone anisotropy 
  in the electron momentum pitch angle. 
The results are shown in Figure~\ref{fig:beam_pl}--\ref{fig:loss_cone_kappa}.
In all the anisotropic cases considered,
the degree of linear polarisation is still
    well described by 
    $\Pi_{\rm L,gen} = (\alpha_\nu+1)/(\alpha_\nu+5/3)$
    at the high frequency end, 
    but deviates at lower frequencies.
The transition frequency that separates these two regimes is different for each case.
This could be explained by the nature of beamed radiation
    of synchrotron radiation.
In general, the synchrotron radiation observed is predominantly contributed by a narrow cone of electrons that are traveling close to the direction of the observer, i.e.~$\tilde{\alpha} \approx\theta$.
We call it the emitting cone hereafter. 
The angular size of the emitting cone $\theta_c$ depends on the 
    electron energy and the observing frequency,
    given by \citep{Yang2018ApJ}: 
\begin{equation} \label{eq:theta_c}
    \theta_c(\gamma, \nu) \sim \frac{1}{\gamma} 
    \left( \frac{2\nu_{\rm c}}{\nu} \right)^{1/3} \ .
\end{equation}
From Eq.~\ref{eq:theta_c}, it can be seen that the emitting cone is smaller for higher frequencies.
As long as the pitch-angle distribution function is fairly constant 
    within the emitting cone, 
    whether the pitch-angle distribution function is isotropic outside
    the cone does not significantly affect the total and polarised flux at that frequency.
When going to lower frequencies,
    the emitting cone becomes larger,
    and the anisotropy, if exist, will impact the emission spectrum, 
    causing the spectrum to differ from an isotropic case,
    even when the energy distributions 
    are the same.
As the formula $\Pi_{\rm L,gen}$
    works well for isotropic cases
    (see Section~\ref{sec:isotropic}),
    it should work well for anisotropic cases
    in the high frequency but not in the low frequency range for the above reason.
Unfortunately, 
    the transition frequency cannot be easily worked out or estimated
    as it depends complexly on the energy distribution function,
    the pitch-angle distribution function, and the viewing angle.
Below we discuss the results of the anisotropic cases in details,
    and show how the results agree with the above explanation.

%
\subsubsection{Beamed distribution}  

Figure~\ref{fig:beam_pl} 
   shows the specific flux $F_\nu(\nu)$
   and degree of linear polarisation 
   $\Pi_{\rm L}$ of synchrotron radiation for electrons that follow a power-law energy distribution 
   with a beamed pitch-angle distribution.
In the case that the electrons beam towards the magnetic field direction ($\tilde{\alpha}_0 = 0$),
    even when the observer is looking right into the beam ($\theta = 0$),
    the observed flux is weak, especially in the high frequency range.
This is because the synchrotron power of an electron scales with $\sin \tilde{\alpha}$.
There is no radiation when the electron motion aligns the 
    magnetic field perfectly, and thus the observed flux is weak when $\theta = 0$.
The low-frequency emission is less affected by this factor because it is contributed by a 
    wider cone of electrons that has a significant non-zero pitch angle.
When the observer gradually looks away from the electron beam,
    two countering factors start to compete with each other.
On one hand, the synchrotron power is stronger for
    larger $\tilde{\alpha}$, and therefore the observed flux is expected to increase.
On the other hand, there are fewer and fewer electrons
    that point towards the observer when looking away from the beam.
The result of these two competing factors is that 
    the observed flux first increases with the viewing angle $\theta$, then gradually decreases after peaking at some sweet spot.
For example, for $\tilde{\alpha}_0 = 0$ and $\sigma_\alpha = 10^{-2}$,
    the observed flux increases from $\theta = 0$ 
    to $\theta = 10\sigma_\alpha$, and gradually 
    drops for $\theta \gtrsim 10\sigma_\alpha$.
Lastly, as pointed out in the previous section,
    the generalised PD formula agrees well with the actual values at 
    high frequency, but deviates at lower frequency.
For electron beam with $\tilde{\alpha}_0 = 0$,
    the transition frequency appears to be insensitive to the electron beam width
    and the viewing angle.

When $\tilde{\alpha}_0 = \pi/2$,
    the observed flux is the strongest when the observer looks directly into the beam ($\theta = \tilde{\alpha}_0$),
    and gradually decreases when the observer looks away from the electron beam.
The transition frequency is insensitive to the viewing angle, but 
    is lower for a larger electron beam. 
This is because a beam with larger width is closer to an isotropic distribution
    for a wider range of pitch angles, making the PD formula $\Pi_{\rm L,gen}$
    valid for a larger frequency range.

Figure~\ref{fig:beam_kappa} is the same as Figure~\ref{fig:beam_pl}
    except that the electron energy distribution follows a 
    kappa distribution instead.
The results are similar overall and 
    almost identical in the high frequency range.
This is due to the property that a kappa energy distribution recovers a 
    power law when the energy is high.
The flux density and PD at low frequencies are slightly different but all discussion on
    the power-law distribution applies to the kappa distribution as well.

%
\subsubsection{Loss-cone distribution} 

\begin{figure*}
    \centering
    \includegraphics[width=0.33\linewidth]{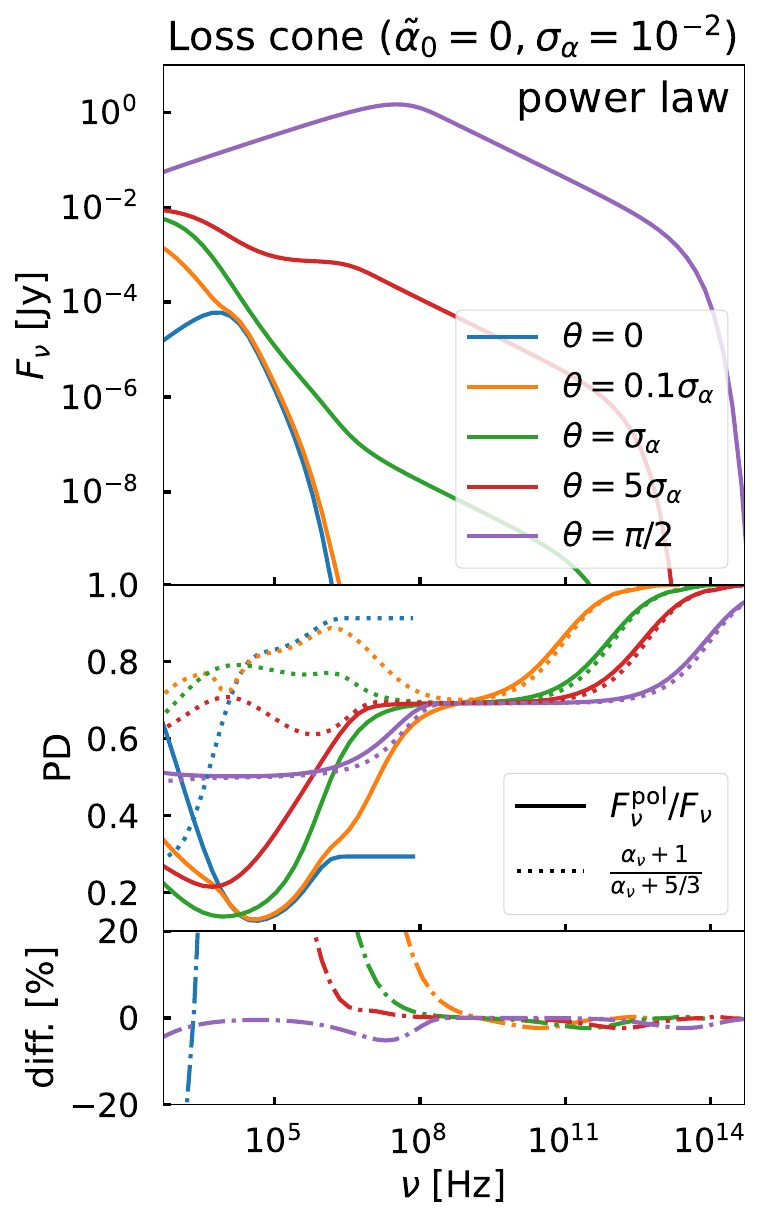}
    \includegraphics[width=0.33\linewidth]{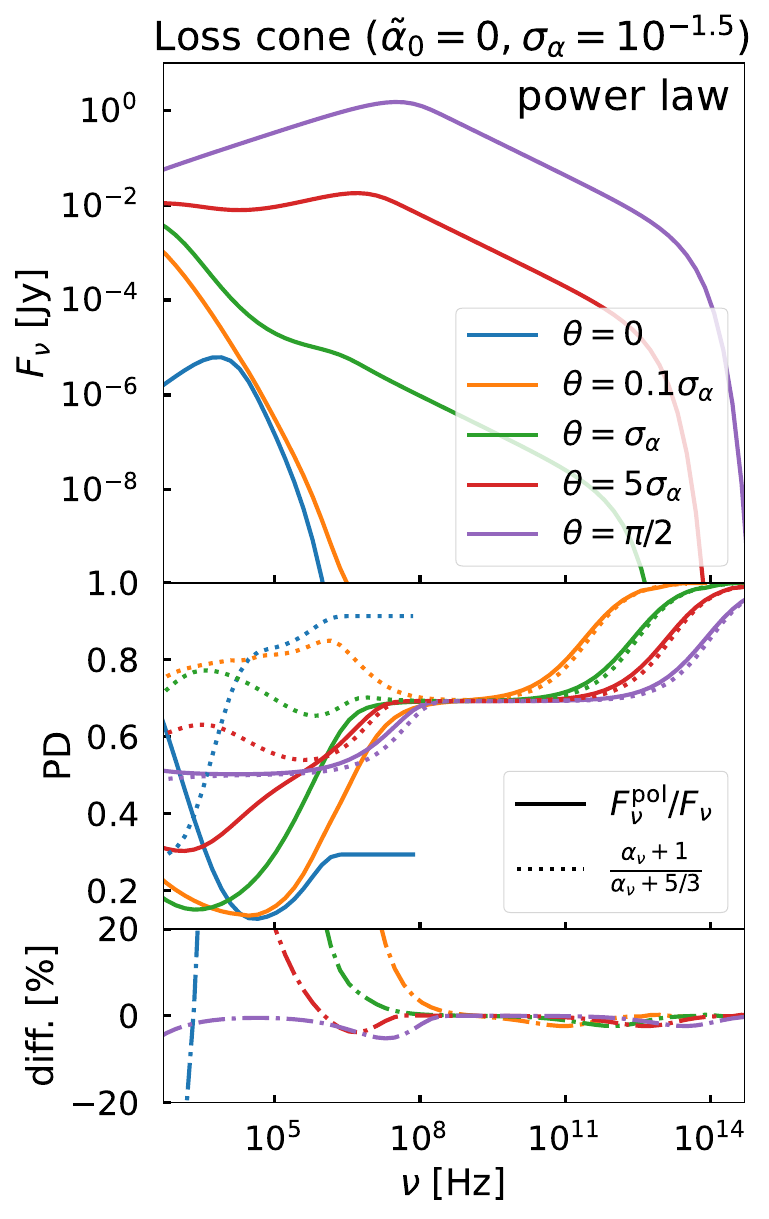}
    \includegraphics[width=0.33\linewidth]{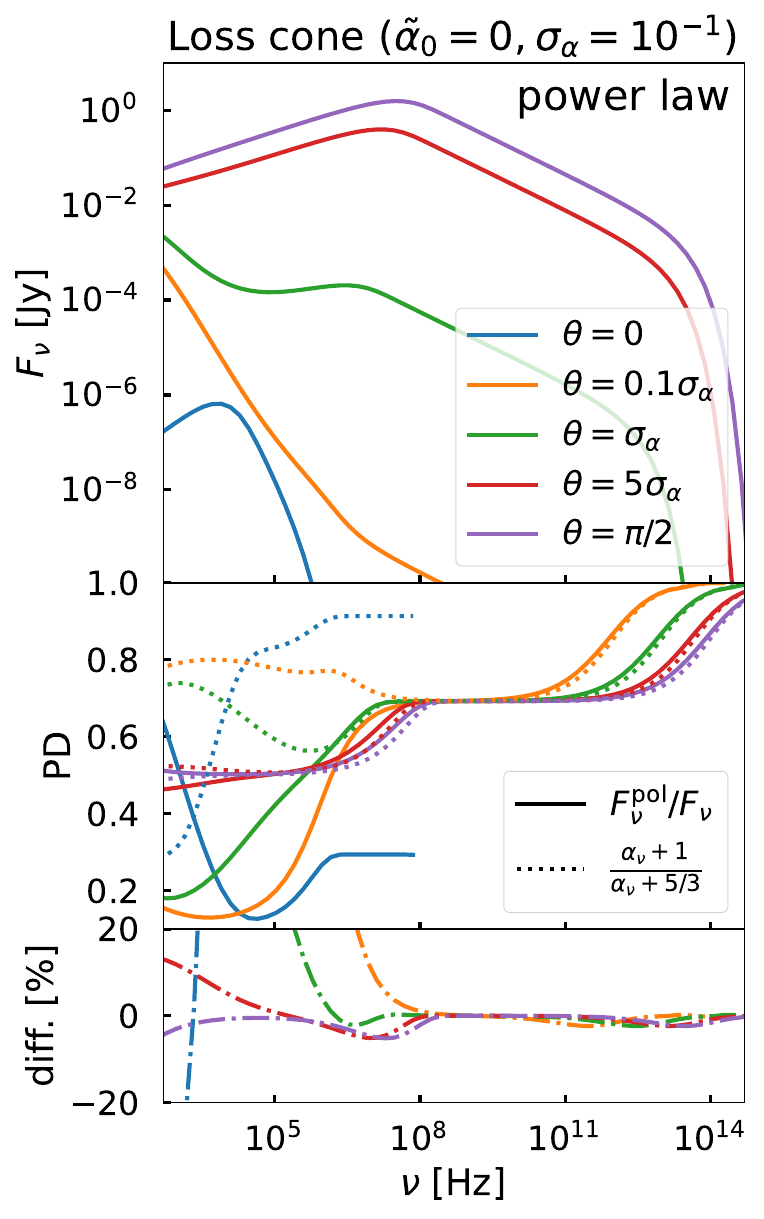}
    
    \includegraphics[width=0.33\linewidth]{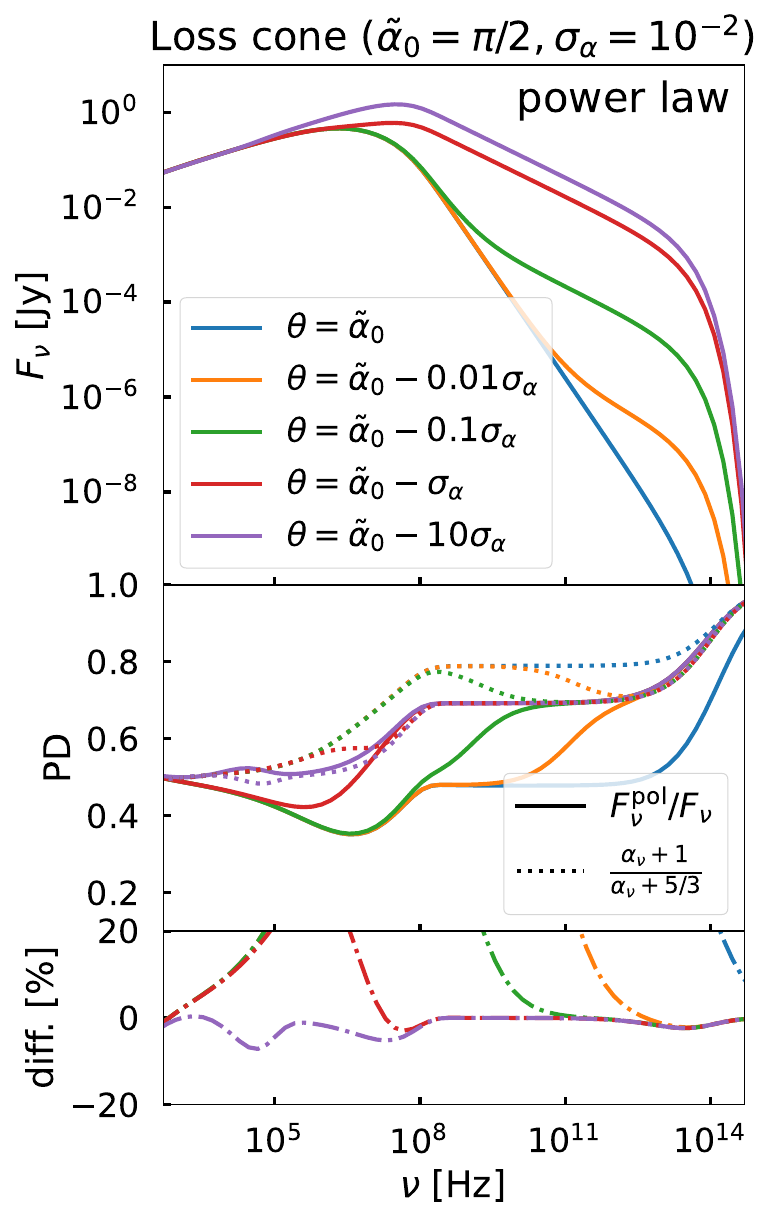}
    \includegraphics[width=0.33\linewidth]{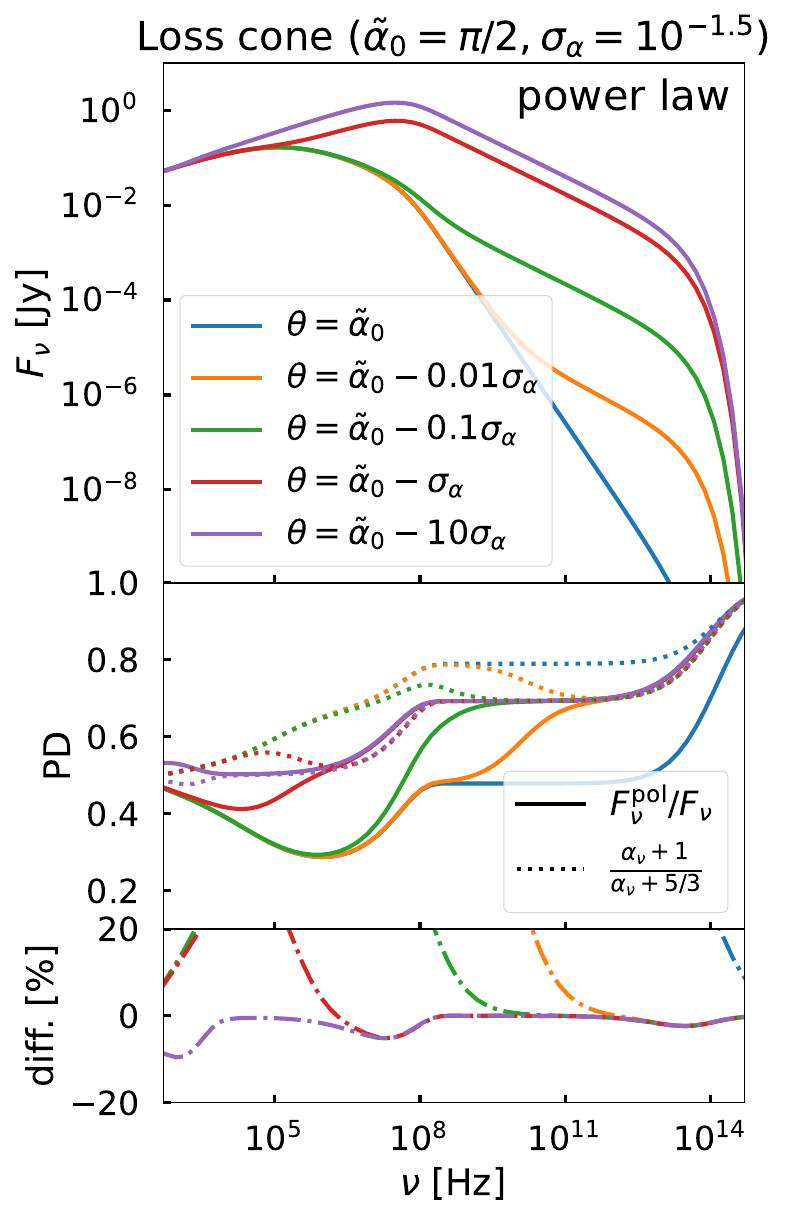}
    \includegraphics[width=0.33\linewidth]{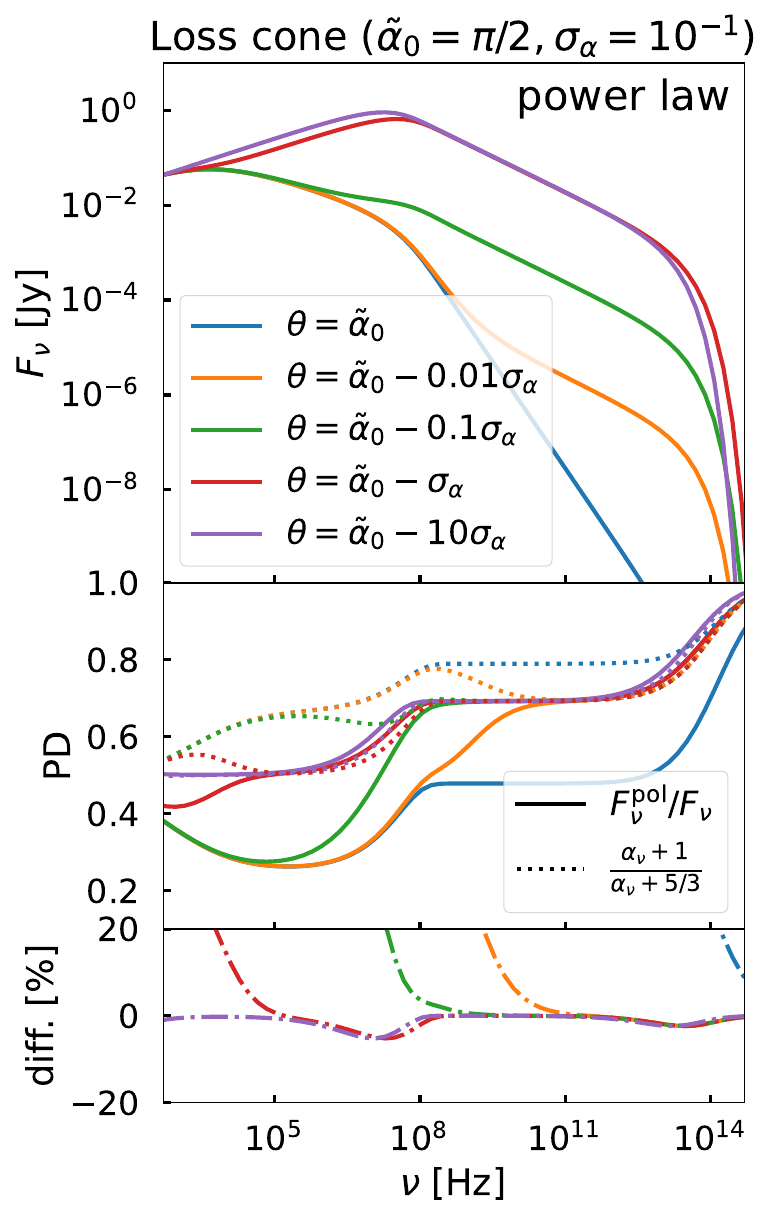}
    \caption{Synchrotron spectra of the loss-cone model with different values of the parameters, i.e.
    the loss-cone direction $\tilde{\alpha}_0$, 
    the opening angle of the loss cone $\sigma_\alpha$,
    and the viewing angle $\theta$.
    For all the calculations, electrons are assumed to follow
    a power-law energy distribution with index $p = 2$, a low-energy cut-off $\gamma=10^2$ and high-energy cut-off $\gamma=10^5$.
    PD is calculated using two methods.
    The solid line is calculated using the full formula (Eq.~\ref{eq:F_omega} and \ref{eq:F_pol_omega}), and the dotted line is calculated from
    the local spectral index $\alpha_\nu$ (Eq.~\ref{eq:pd_gen}).
    The cone width of these models are 
    $\alpha_{\rm HWHM} = 0.15, 0.27, 0.49, 0.012, 0.037, 0.12$ from left to right, from top to bottom.
    Besides, the following parameters are used:
    $N_{e, 0} = 10^{58}$, $B = 1\,$mG, and $D = 1$\,Gpc.}
    \label{fig:loss_cone_pl}
\end{figure*}

\begin{figure*}
    \centering
    \includegraphics[width=0.33\linewidth]{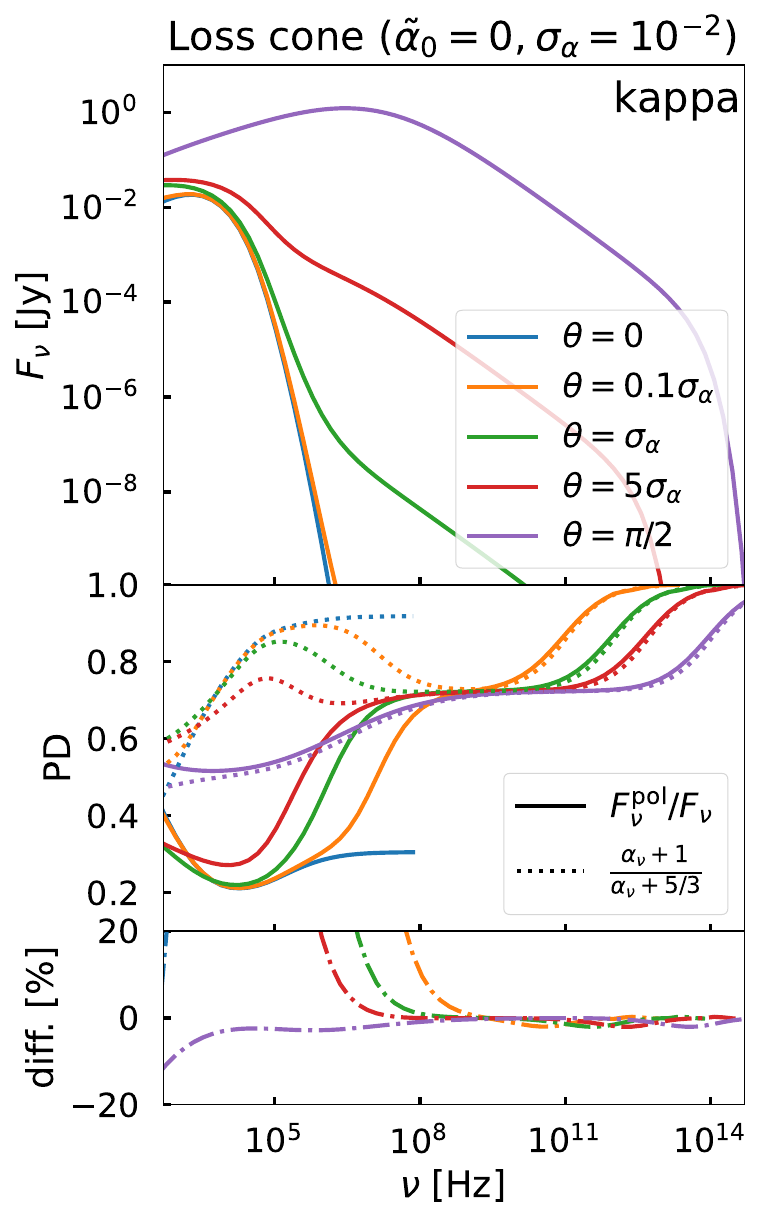}
    \includegraphics[width=0.33\linewidth]{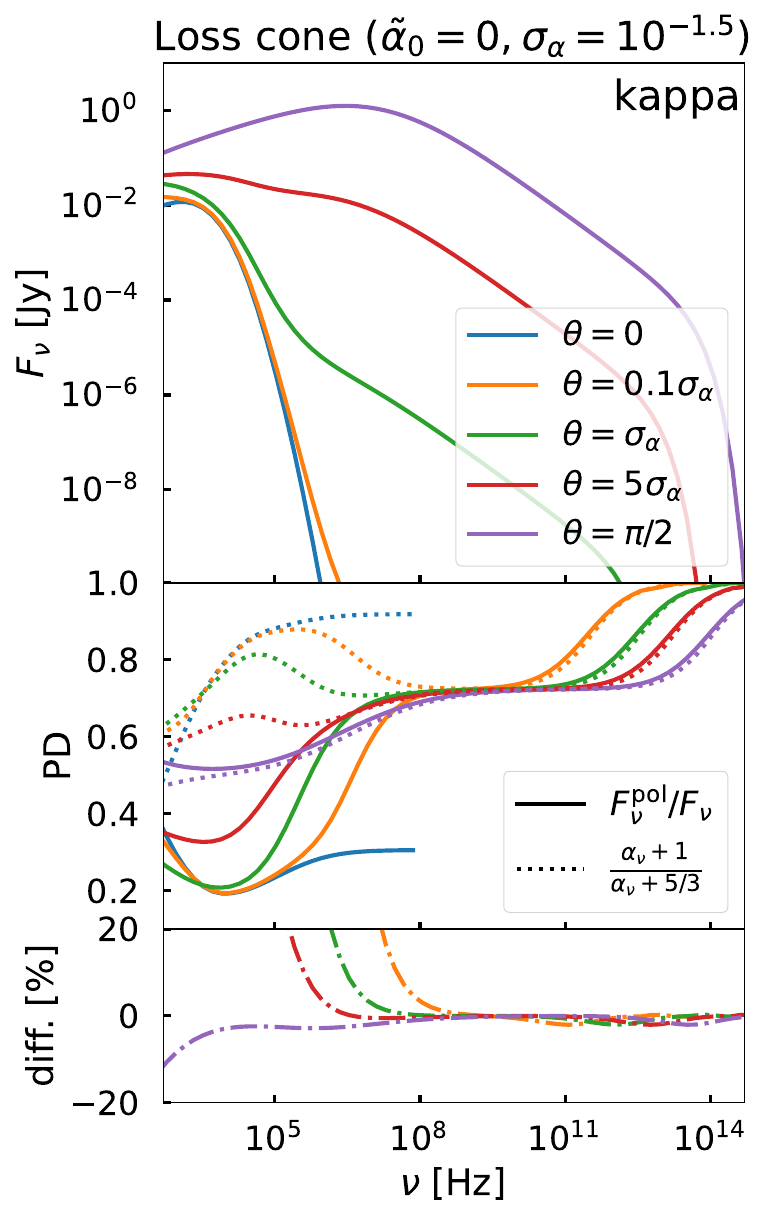}
    \includegraphics[width=0.33\linewidth]{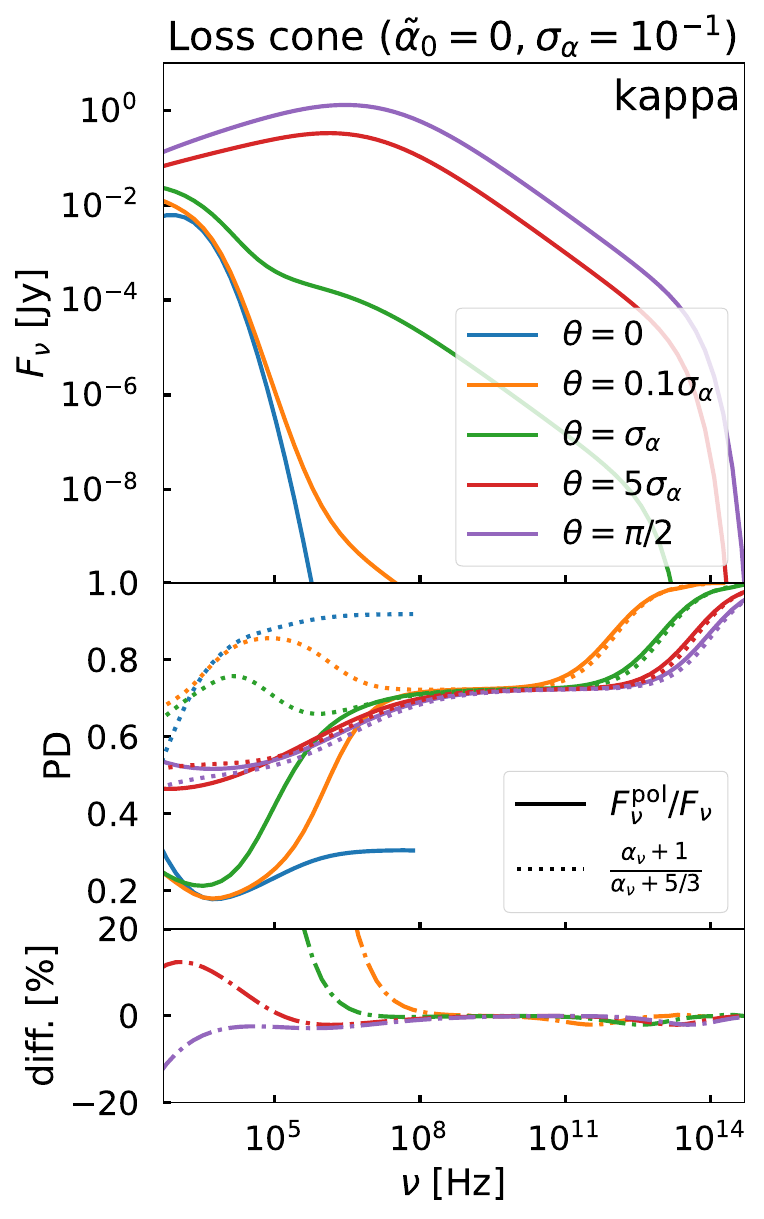}
    
    \includegraphics[width=0.33\linewidth]{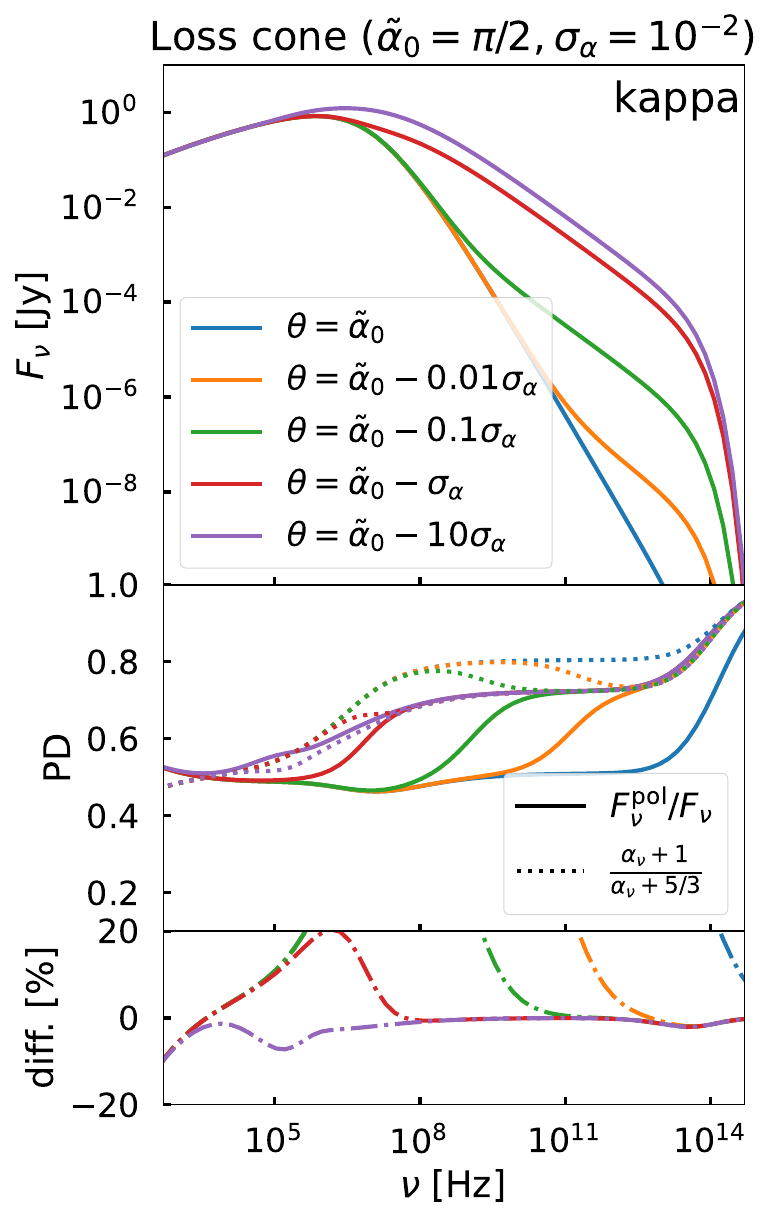}
    \includegraphics[width=0.33\linewidth]{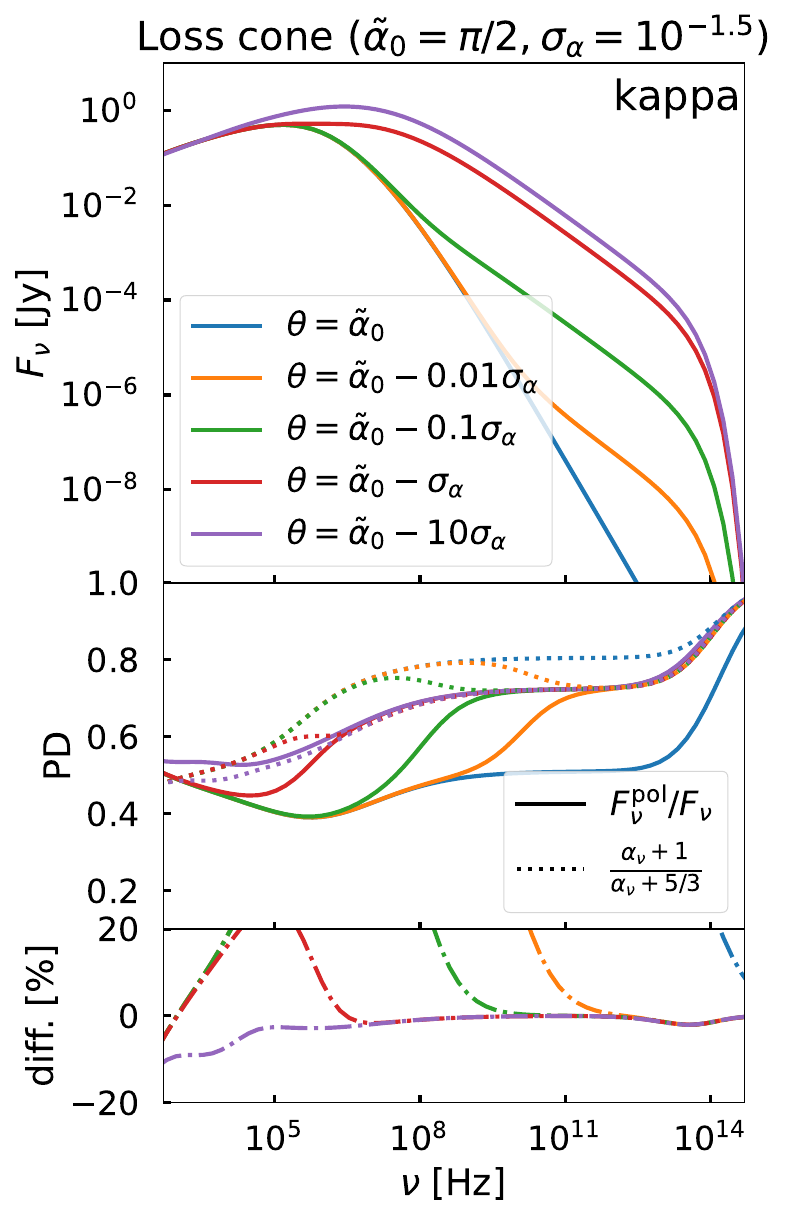}
    \includegraphics[width=0.33\linewidth]{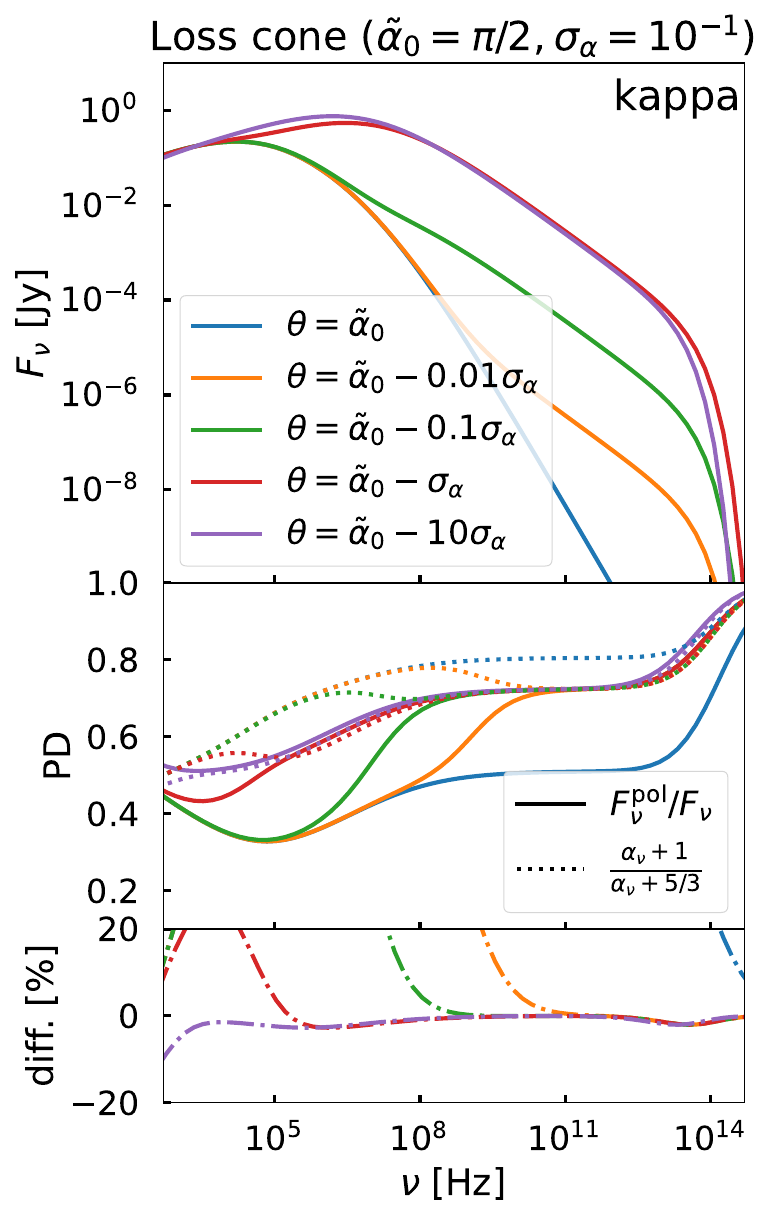}
    \caption{Synchrotron spectra of the loss-cone model with different values of the parameters, i.e.
    the loss-cone direction $\tilde{\alpha}_0$, 
    the opening angle of the loss cone $\sigma_\alpha$,
    and the viewing angle $\theta$.
    For all the calculations, electrons are assumed to follow
    a kappa distribution with $\kappa = 3.5$ 
    and $w = 10$, and a high-energy cut-off $\gamma=10^5$.
    PD is calculated using two methods.
    The solid line is calculated using the full formula (Eq.~\ref{eq:F_omega} and \ref{eq:F_pol_omega}), and the dotted line is calculated from
    the local spectral index $\alpha_\nu$ (Eq.~\ref{eq:pd_gen}).
    The cone width of these models are 
    $\alpha_{\rm HWHM} = 0.15, 0.27, 0.49, 0.012, 0.037, 0.12$ from left to right, from top to bottom.
    Besides, the following parameters are used:
    $N_{e, 0} = 10^{58}$, $B = 1\,$mG, and $D = 1$\,Gpc.}
    \label{fig:loss_cone_kappa}
\end{figure*}

Figure~\ref{fig:loss_cone_pl} 
   shows the specific flux $F_\nu(\nu)$
   and degree of linear polarisation 
   $\Pi_{\rm L}$ of synchrotron radiation for electrons that follow a power-law energy distribution 
   with a loss-cone pitch-angle distribution.
For any values of $\tilde{\alpha}_0$ and $\sigma_\alpha$, the observed flux is the lowest when looking directly into the loss cone ($\theta = \tilde{\alpha}_0$).
Also, when $\theta = \tilde{\alpha}_0$, 
    the observed flux is lower for a larger loss cone.
These two outcomes showcase that the emission is mainly contributed by particles traveling close to the direction of the observer.
Unlike the beamed distribution, the observed flux always increases
    with the viewing angle for the loss-cone distribution when $\tilde{\alpha}_0 = 0$.
This is caused by the increase in particle number for larger pitch angle $\tilde{\alpha}$ and the scaling of synchrotron power with $\sin\tilde{\alpha}$.
When $\tilde{\alpha}_0 = 0$, the PD formula $\Pi_{\rm L,gen}$ agrees well with the actual number in the high-frequency range.
The transition frequency is lower for larger $\theta$,
    and $\Pi_{\rm L,gen}$ almost matches the entire frequency range when $\theta = \pi/2$.
This is due to the fact that a loss-cone distribution is almost isotropic outside the small loss cone.
When $\tilde{\alpha}_0= 0 $ and $\theta = \pi/2$, the spectral energy distribution of the loss-cone distribution is only slightly different from an isotropic distribution.

When $\tilde{\alpha}_0 = \pi/2$,
$\Pi_{\rm L,gen}$ gives bad estimation of the PD for some of the cases.
For example, when the observer looks directly into the loss cone ($\theta = \tilde{\alpha}_0$), $\Pi_{\rm L,gen}$ is invalid for the entire frequency range.
Anisotropy is the main cause of this discrepancy but the interpretation is different depending on the observing frequency considered.
In the low frequency range, the emitting cone is so large that the anisotropy is naturally manifested in the spectrum, similar to the reason discussed in the beamed distribution case.
In the high frequency range, although the emitting cone is small and the pitch-angle distribution is fairly constant within it, there lacks particles inside the emitting cone and the emission mainly comes from particles outside of it.
If the spectrum significantly differs from an isotropic distribution, $\Pi_{\rm L,gen}$ will fail to match the real PD.
When the observer gradually looks away from the loss cone, 
$\Pi_{\rm L,gen}$ starts to agree with the actual PD again in the high frequency range,
and is almost identical when the viewing direction is far away from the loss cone,
e.g., when $\tilde{\alpha}_0 = \pi/2$, 
$\sigma_\alpha = 10^{-2}$, and 
$\theta = \tilde{\alpha}_0 - 10\sigma_\alpha$.
This is again due to the fact that a loss-cone distribution is almost isotropic outside the small loss cone.

In Figure~\ref{fig:loss_cone_kappa}, the results are calculated using a kappa energy distribution.
Overall, the results are similar to the power-law case (Figure~\ref{fig:loss_cone_pl}), with the high-frequency result being nearly indistinguishable between the two.
This similarity arises because the kappa distribution approaches a power-law form at high energies. At lower frequencies, there are minor differences in the flux density and PD, but all conclusions drawn for the power-law distribution remain applicable to the kappa distribution as well.

%
\section{Discussion} 
\label{sec:Discussion} 

%
\subsection{Inhomogeneities in the emission region 
  and propagation effects} 
\label{sec:structure_propagation}

The PD calculations presented above,  
  which have put focus on 
  the properties of emission particles, 
  have adopted the following assumptions: 
  (i) the magnetic field in the emission region 
      is uniform structurally 
      and has a well defined value, and 
  (ii) the overall particle density  
      in the emission region is uniform.  
These assumptions allow us to compute 
   the spectrum and the PD 
   of synchrotron radiation 
   from an ensemble of charged particles 
   unambiguously.  
We refer the PD obtained in this idealised setting,  
  as intrinsic PD. 

We now discuss the effects on the PD 
  of synchrotron radiation 
  and what we would expect for the measured PD 
  when the two assumptions are relaxed.   
We first examine assumption (i), 
  keeping a uniform particle properties 
  and density across the emission region. 
This assumption is generally valid,  
   for computing the local emission from 
   within an astrophysical system, 
   when its linear size $L > l_{\rm mag} > r_{\rm gyr}$, 
   where $r_{\rm gyr}$ is the gyration radius 
   of the emitting charged particles  
   and $l_{\rm mag}$ is the length scale 
   over which the magnetic field 
   in the emission region varies.   
The measured PD will be lower than 
    the intrinsic PD, 
    for a given characteristic strength 
    of the magnetic field in the emission region,   
    as emission from different locations 
    have different orientations of polarisation which could partially cancel out each other.
Relaxing assumption (ii) has no effect on the PD 
  of the total synchrotron radiation,  
  if the magnetic field is uniform 
  across the emission region 
  and the condition $L \gg r_{\rm gyr}$ holds 
  and the total number of charged emitting particles 
  is fixed, 
  as PD is an intensive quantity. 
The PD could, however, be significantly affected,  
  as shown in Section ~\ref{sec:anisotropic}, 
  if the particle distribution  
  deviates from isotropy, in the momentum space.  
  
Polarised radiation can be modified 
  in propagation, 
  through Faraday rotation and conversion 
  \citep[see][]{Pacholczyk1967ApJ,Sazonov1968R&QE} 
  and also polarisation dependent absorption 
  \citep[see][]{Pacholczyk1967ApJ,Jones1977}. 
Faraday leads to the conversion between 
  two Stokes parameters, $U$ and $Q$,   
  along a single ray, which alters the polarisation angle 
  but does not affect the PD.  
For an extended synchrotron source 
  with inhomogeneous magnetic-field structure, 
  the total emission 
  is the sum of contributions of 
  all rays reaching that observer.  
These rays have different amount of Faraday rotation 
  in their propagation 
  (e.g. when they pass through 
  a turbulent medium) 
  in addition to different initial Stokes $U$ and $Q$.   
The measured PD of the emission 
   contributed by a bunch 
   of spatial unresolved rays  
   is therefore always lower than the intrinsic PD 
   of the same set of rays 
   if both the emission region 
   and the medium between it and observer 
   have uniform magnetic field  
   and particle density.  
This effect 
  would contribute to  
  the so-called  
  beam depolarisation in external Faraday dispersion.

PD, only accounting for linear polarisation, 
  can be modified   
  in the presence of inter-conversion between 
  linear polarisation and circular polarisation modes 
  \citep[see e.g.][]{Pacholczyk1977OISNP,Melrose1991}   
  and mode-dependent absorption   
    \citep[see][]{Pacholczyk1967ApJ,Jones1977}
    in propagation.   
It could also be modified 
   by boundary effects arisen 
   \citep[see e.g.][]{Hecht2001opt4}
    for polarised radiation traversing a composite medium, 
    consisting of components 
    with substantial gradients 
    in their refractive indices. 
This effect is well known in laboratory physics  
   but is seldom explored 
   in astrophysical settings\footnote{For instance,    
   when low-frequency synchrotron radiation 
   traverses a foreground current sheet,   
   such as those associated 
   with a collisionless astrophysical shocks, 
   the PD of the synchrotron radiation 
   will inevitably be affected. 
  Observations have shown that AGN
   could have substantial 
   radio synchrotron radiation as low as $100~{\rm MHz}$ 
   \citep[see e.g.][]{Calistro-Rivera2017MNRAS}. 
   The study of \cite{Singh2013A&A} on 
     Seyfert galaxies 
     shows that a good fraction of these sources 
     have similar values of spectral indices measured between 240\,MHz, 610\,MHz, 1.4\,GHz, and 5\,GHz,
     hinting that the radio emission from AGN
     could extend to $100\;\!{\rm MHz}$, 
     provided the radiation 
       is not fully suppressed 
       by the plasma-frequency cut-off 
       of the line-of-sight medium.}.

%
\subsection{Practical applications in observations} 

Eq.~\ref{eq:pd_pl}, 
  together with the formula 
  for rotation measure,\footnote{In a magnetised plasma with 
    a spatially varying magnetic field ${\boldsymbol B}$, 
   an expression for the rotation measure accounting for both thermal and non-thermal electrons is given by  
\begin{align}  
  {\mathcal R}(s) = \frac{e^3}{2\pi {m_{\rm e}}^2 c^4} 
    \int^s_{s_0} 
      {\rm d}s' \  n_{\rm e}(s') \Theta(s') 
        B_{\parallel}(s')   
\end{align} 
  \citep[see][]{On2019MNRAS}, 
 where the electrons 
 have an isotropic momentum distribution and 
 a total number density $n_{\rm e}$. 
The variable 
  $\Theta (s) = 1 - \Upsilon(s)[1- \zeta(p,\gamma_{\rm i})]$ 
  is a factor weighing the contribution 
   of thermal and non-thermal electrons, given the fraction of non-thermal electron 
   $\Upsilon(s) = n_{\rm e, nt}(s)/n_{\rm e}(s)$. 
   For non-thermal electrons 
   with a power-law energy distribution 
   of an index $p$, 
\begin{align} 
  \zeta(p,\gamma_{\rm i}) 
    = \frac{(p-1)(p+2)}{(p+1)} 
    \left(\frac{\ln \gamma_{\rm i}}{{\gamma_{\rm i}}^2}\right) \ 
\end{align} 
  (for $p> 1$), where $\gamma_{\rm i}$ 
  is the Lorentz factor of low-energy cut-off 
  of the non-thermal electrons.} 
  which is defined as 
 ${\mathcal R} \equiv (\Delta \varphi)\;\! \lambda^{-2}$  
     (where $\lambda$ 
     is the wavelength of the radiation,and  
     $\varphi= (1/2) \tan^{-1}(U/Q)$)  
 has been a work-horse in astrophysics 
  for the studies of emission of synchrotron sources 
  and the line-of-sight modification 
  of the polarised synchrotron radiation. 
An ``algorithm'' as follow is often used 
  for the interpretations of polarised radiation,    
  in particular in the radio wavelengths 
  as the emission mechanism of 
  many non-thermal source is of synchrotron origin:  
(1) Measure the total fluxes, 
   polarised fluxes, and polarisation angle   
   of the radiation from sources 
   at several wavebands. 
(ii) Make plots of fluxes, 
   polarised fluxes and polarisation angle 
   as function of wavelength for sources. 
(iii) Deduce $p$, PD and ${\mathcal R}$ from the plots.  
(iv) Apply these information 
   to set constraints 
   on the physical and dynamical processes in the source  
   and on the properties of magnetised medium 
   between the sources and the observers. 
This algorithm is often based on an implicit assumption 
  that the momentum distribution  
    of the emitting particles is isotropic 
  and they gyrate in a magnetic field 
  in radii much smaller than the 
  length scale over which 
  the magnetic field varies 
  in the emission region  
  and in the medium the radiation traverses. 

Modification of PD can be interpreted 
  as the presence of structural inhomogeneities 
  in the magnetic fields 
  (e.g. global spatial non-uniformity
  or tangled field lines)    
  or abnormal number density distribution of charged particles 
   (e.g. large density gradients, jumps or turbulence), 
   which are presumably electrons and/or positrons.    
An extended power law with a positive spectral index $\alpha$ at high frequencies 
  is attributed to optically thin emission. 
The power-law index $\alpha$ 
  is an indicator of whether or not the charged particles 
  is freshly accelerated (with $\alpha \approx 0.7$) 
  or has aged (with $\alpha \sim 1$ or higher). 
A spectral turn-over at low frequencies 
  signals an optically thin to optically thick transition, 
  and the frequency at which the radiation peaks 
  is determined 
  by the strength of the magnetic field 
  in the emission region (self-absorption)
  or in the medium 
  between the source and the observer (external absorption). 
  
Shock acceleration processes generally leads to 
    $p \approx 2.3 - 2.5$ 
  \citep[see][]{Bell1978MNRAS,Blandford1978ApJ,Kirk2000ApJ},
  which corresponds to an intrinsic PD of $\sim$70\%
  (using Eq.~\ref{eq:pd_pl}).
A 70\% PD is commonly considered as the canonical upper limit
    for synchrotron-emitting astrophysical objects,
    as the observed values are expected to be lower than
    that due to structural inhomogeneity 
  and propagation effects 
  as described in Section \ref{sec:structure_propagation}.
For point sources, 
  the PD of individuals 
  could be contaminated 
  by background and foreground 
  emission from line-of-sight diffuse media,  
  when they are magnetised 
  and their emission is non-negligible 
  compared to sources 
  \citep[see][]{On2025PASA}. 

In the standard algorithm 
  $\alpha$ is determined  
  by fitting an extended flux spectrum  
  crossing over a substantial range 
    of wavelengths (or frequencies).    
It is then used 
  to obtain $p$ and PD. 
Although it may sound trivial, 
  the reality is that it is not always possible to obtain 
  a sufficient coverage in wavelength (or frequency) 
  in observations to confidently determine $\alpha$,
  even when 
  the emission spectrum is actually 
  an extended power law.    
Moreover, there is no guarantee 
  that the emission spectrum is a power law.  
Thus, using the value of $\alpha$ obtained from the fit to calculate $p$ and the corresponding PD becomes a meaningless exercise,
if the emitting charged particles do not have a power-law energy distribution.
Adopting the PD derived from this fit
  will give misleading interpretation 
  at best 
  and may even lead to incorrect scenarios 
  or physical models. 
 
The findings of our study imply 
  that we can construct more robust algorithms 
  that are reliable beyond an extended power-law spectrum 
  and are applicable when the observations 
  lack broad multi-wavelength (multi-frequency) coverage. 
What we need is to determine the local spectral index 
  $\alpha_\nu$, 
  instead of the global spectral index $\alpha$.   
The index $\alpha_\nu$ is a local quantity, 
  which is well defined at the frequency $\nu$ 
  if the spectral segment over $\nu$ is sufficiently smooth, unlike $\alpha$, which is ill defined 
  when the spectrum is not a power law.  
As shown in Section~\ref{sec:results},
  Eq.~\ref{eq:pd_gen}, which makes use of 
  $\alpha_\nu$, 
  gives PD that matches the analytical value surprisingly well.
Eq.~\ref{eq:pd_gen} 
  is therefore the more general and appropriate 
  expression for intrinsic PD,
  provided that the momentum distribution 
  of the emitting charged particles is sufficiently isotropic.

\subsection{Remarks on momentum isotropy of charged particles} 

We have demonstrated  
  the versatility of the generalised formula Eq.~\ref{eq:pd_gen} 
  for the analysis of polarised synchrotron radiation 
  and inference of the energy distribution of 
  the emitting charged particle, 
  when the radiation 
  does not have an extended power-law spectral segments. 
The applicability of Eq.~\ref{eq:pd_gen} 
  requires that 
  the momentum distribution of emitting charged particles 
  is isotropic. 
As shown in our calculations the spectral properties 
    of the emission from particles 
    with anisotropic momentum distribution 
  can substantially differ to those 
    of the emission from particles 
    with isotropic momentum distribution.
Recent studies highlighted the difference in the spectral properties 
  of the synchrotron radiation 
  for electrons with isotropic and anisotropic momentum \citep{Yang2018ApJ, Comisso2020ApJ} (see also an earlier 
  study by \cite{Robinson1985PPCF} 
   for a more general plasma and space physics context).
We extend their work 
  to include polarisation.  
Unsurprisingly, 
  the polarisation properties 
  of synchrotron radiation 
  from charged particles with 
  isotropic and anisotropic particles 
  can differ substantially, 
  which can be seen in the cases shown 
  in Section~\ref{sec:anisotropic}. 

In modeling of radiation from astrophysical sources,  
  the momentum distribution of the emitting particles 
  is often assumed to be isotropic, for simplicity or 
  for avoiding unmanageable complications  
  especially when we lack information 
  about the exact microscopic properties 
  regarding the emission charged particle distribution. 
In certain situations, 
  the assumption of isotropic momentum distribution 
  is reasonable, 
  e.g. when the charged particles are scattered stochastically 
    by turbulent magnetic fields. 
In many astrophysical situations, 
 the momentum distribution of the charged particles 
 would have some degree of anisotropy. 
As pointed out in \cite{Yang2018ApJ},   
  momentum distribution anisotropy 
  can arise from anisotropic accelerations 
  such as in relativistic shock waves 
  with a spatially ordered magnetic field. 
Phenomena such as magnetic reconnection 
  and magnetic mirroring 
  can also cause anisotropy in the particle momentum distribution 
  \citep[see][]{Yang2018ApJ,Comisso2019ApJ,Lazarian2021ApJ,Xu2023ApJ}.
Momentum anisotropy can also arise 
  from strong radiative cooling \citep{Comisso2021PhRvL}
  and from propagation 
  of relativistic charged particles 
  in certain electro-magnetic field configurations 
  (cf. the developing of loss-cone and horseshoe
  momentum distribution 
  as those in the cyclotron maser sources, 
  see \cite{Melrose1982ApJ}, \cite{Ergun2000ApJ} 
  and \cite{Willes2004MNRAS} for example).  
In addition, anisotropic electrons provide explanations to some of the astrophysical phenomenon, such as the limb-brightening of AGN jets \citep{Tsunetoe2025ApJ}.

\cite{Comisso2023ApJ} 
  presented polarisation calculations 
  of synchrotron radiation from 
  charged particles with anistropic momentum distribution 
  with a power-law energy spectrum. 
A more general and comprehensive exploration of 
  the effects on polarisation properties 
  of synchrotron radiation 
  from charged anisotropic momentum distribution 
  for various energy spectra 
  is presented in this work. 
Similar to \cite{Comisso2023ApJ}, 
  we have found that the polarisation properties 
  are strongly affected by the anisotropic particle momentum, 
  and therefore the canonical upper limit of roughly 70\% 
  for optically thin synchrotron radiation 
  would deem to be inapplicable 
  in many astrophysical situations 
  where isotropic momentum  of the charged particles 
  are not guaranteed \cite[see][]{Yang2018ApJ, Comisso2023ApJ}.    
It is important to be cautious 
  on the conditions for the applicability of certain equations 
  in relating the PD and spectral properties 
  in synchrotron sources in astrophysics 
  and in imposing the canonical limits of PD 
  derived from the assumption 
  of isotropic momentum distribution 
  and an extended power-law energy spectrum 
  of the emitting charged particles. 

\subsection{Synchrotron radiation 
  emitted by other charged particles}  

  We have presented the calculations for the spectro-polarisation of 
 synchrotron radiation from electrons with 
 various energy distributions, 
 relaxing the restriction that the electrons have an isotropic momentum distribution. 
In astrophysical systems, relativistic electrons 
  are not the only emitters of polarised synchrotron radiation. 
Pairs are common in relativistic astrophysical plasmas  
  as the energetic particles would exceed the 1-MeV threshold.  
The results that we have obtained from synchrotron radiation 
  for electrons are directly applicable to synchrotron radiation for 
  positrons, as all the expressions have no dependence on 
  the odd power of particle charges.  
In violent astrophysical sources, 
  e.g. gamma-ray bursts and relativistic jets/outflows in some AGN, 
  heavier charged particles, such as muons and protons, could be present, 
  and these particles are also emitter of synchrotron radiation. 
When there are substantial amounts of these particles, 
  their radiation would have noticeable contribution 
  to the spectral energy distributions of the sources 
  \citep[see e.g.,][]{Cerruti2015MNRAS, Sahakyan2023MNRAS, Xue2023PhRvD, Zhang2025JHEA}. 

We note that with appropriate scaling, 
  the results obtained for electrons and positrons 
  can be generalised for the other charged particles. 
In the classical theory of radiation, 
  the essential parameters in the emitting particles, denoted as ${\rm X}$ 
  are their mass $m_{\rm X}$ and their electric charge $q_{\rm X}$. 
For synchrotron process, 
  the frequency of the radiation scales with the charge-to-mass ratio 
  $(q_{\rm X}/m_{\rm X})$, 
  of the particles, when their Lorentz factor is fixed. 
The power of radiation scales with the Thomson cross section, 
  which is proportional to $({q_{\rm X}}^2/m_{\rm X})^2$, 
  of the particles, 
  and hence the scaling factor is $({q_{\rm X}}^2/m_{\rm X})^2$.   
In a heuristic argument, 
  for charged particles, with either isotropic or anisotropic momentum pitch-angle distributions,     
  the expressions that we have obtained 
  in Section~\ref{subsec:basic} could be expressed 
  in a generalised form, in term of $q_{\rm X}$ and $m_{\rm X}$, 
  with electrons as the special case, 
  where $q_{\rm X} = e$ and $m_{\rm X} = m_{\rm e}$.  
The generalised expressions for arbitrary charged particles X 
  are the same expressions as those shown in Section~\ref{subsec:basic} 
  but with $e$ replaced by $q_{\rm X}$ 
  and $m_e$ by $m_{\rm X}$.
  
The generalised formula $\Pi_{\rm L,gen} = (\alpha_\nu +1)/(\alpha_\nu +5/3)$ 
  (Eq.~\ref{eq:pd_gen})  
  is also applicable to all charged particles with an isotropic pitch-angle distribution. For anisotropic distributions, it is still applicable to the high-frequency range for some cases, but one must be careful how this frequency range shifts according to the charge and mass of the emitting particles. 
The most important implication is that,
    if the observed radiation is evidently 
    synchrotron radiation,
    the generalised PD formula can be applied regardless of the identity of the emitting particles, given the assumption that the pitch-angle distribution is isotropic.

%
\section{Conclusions}

We have generalised 
  the commonly employed formula  
   $\Pi_{\rm L,pl} = (p+1)/(p+7/3) = (\alpha+1)/(\alpha+5/3)$ 
  for determining the degree of linear polarisation 
  of synchrotron sources in this work.
We show that 
  when the global spectral index $\alpha$ is replaced by $\alpha_\nu$,
  which is locally defined, 
  the generalised PD formula 
  $\Pi_{\rm L,gen} = (\alpha_\nu +1)/(\alpha_\nu +5/3)$
  will give an excellent approximation 
  to the degree of linear polarisation obtained 
  by direct emissivity calculations,  
  provided that the emitting charged particles 
  have an isotropic momentum distribution. 
The new formula that we present 
  provides a reliable tool 
  in the prediction and in the interpretation 
  of the polarisation properties 
  of astrophysical synchrotron sources 
  when observations do not have 
  a broad band coverage in frequencies 
  or in wavelengths. 
We also conduct calculations of linear polarisation 
  of synchrotron radiation 
  from charged particle with anisotropic 
  momentum distribution.
We present results of two special cases:   
  the beamed distribution 
  and the loss-cone distribution. 
Our calculations have shown that 
  both the total and the polarised intensity 
  differ 
  to those of correspondence cases 
  where the charged particles 
  have isotropic momentum.
We note that 
  for anisotropic distributions, 
  the generalised formula that we present  
  may works for a certain frequency range, 
  and is generally inapplicable at the low frequencies.
This indicates that  
  the anisotropy in the momentum distribution 
  of the emitting particles 
  would need to be taken in consideration 
  when interpreting the spectral polarimetric data 
  of astrophysical synchrotron radiation sources.

%
\section*{Acknowledgements} 

We thank Dr Po Kin Leung (CUHK) 
  for valuable discussions 
  on radiative processes in astrophysical systems.
We thank the referee for the useful suggestion.
PCWL is supported by a UCL Graduate Research Scholarship 
  and a UCL Overseas Research Scholarship. 
KJL is supported by a PhD Scholarship 
  from the Vinson and Cissy Chu Foundation 
  and by a UCL MAPS Dean’s Prize.  
YXJY and AKHK are supported by National Science and Technology Council of the Republic of China (Taiwan) through the grant 113-2112-M-007-001.   
KW acknowledges the support 
  from the ANU Distinguished Visitor award 
  and thanks the hospitality of 
  the ANU Research School of Astronomy and Astrophysics 
  and Centre for Gravitational Astrophysics during his visits 
  and thanks the hospitality of 
  the NTHU IoA, where part of 
  this work was conducted, during his visits.  
PCWL, YXJY and KW acknowledge the support from 
  the UCL Cosmoparticle Initiative. 
This research has made use of NASA’s Astrophysics Data System.

%
\section*{Data Availability}

No new data were generated or analysed in support of this research.



\bibliographystyle{mnras}
\bibliography{sync} 




%
\appendix  

%

\section{Isotropic mono-energy electrons} \label{app:mono}

\begin{figure*}
    \centering
    \includegraphics[width=0.43\linewidth]{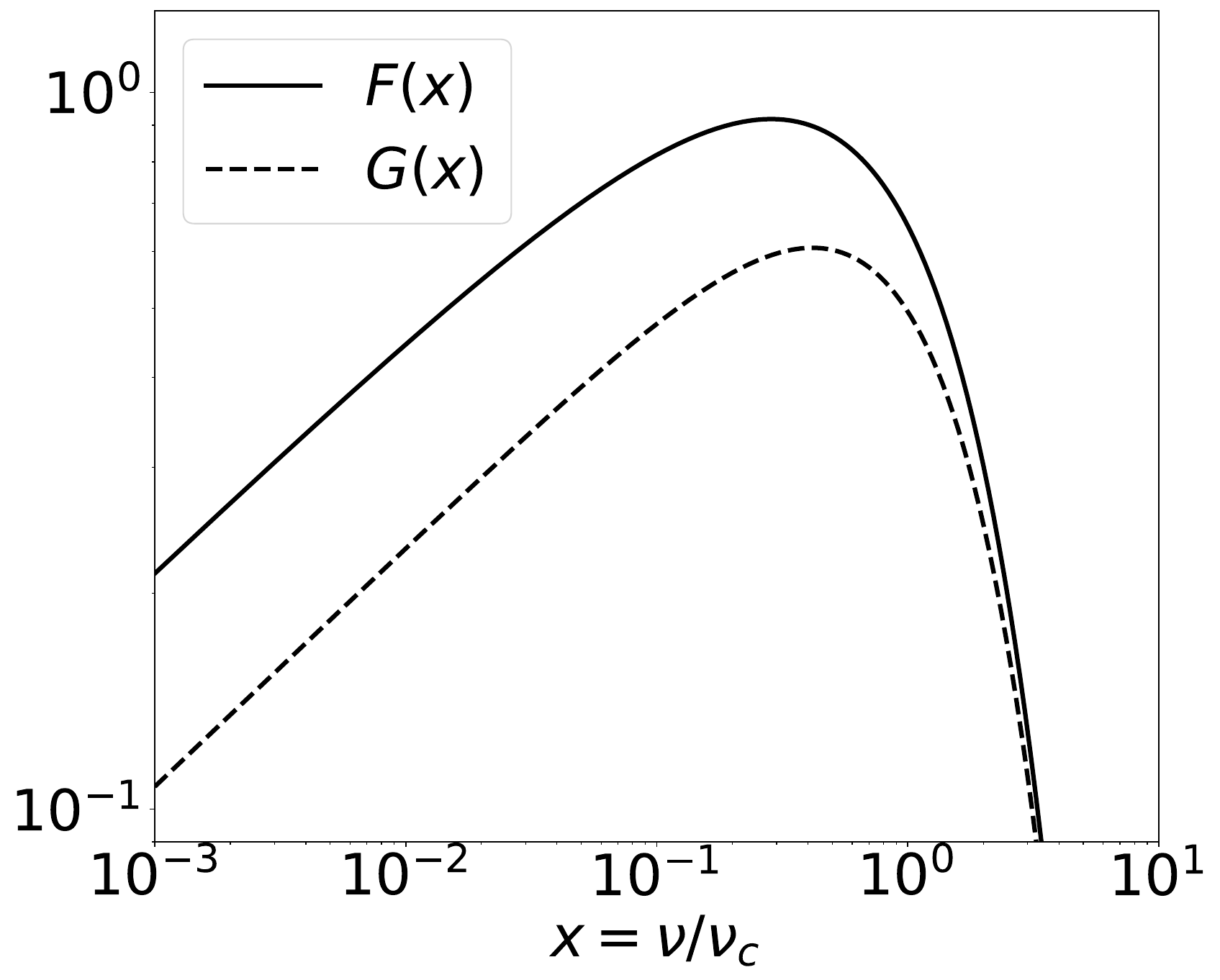}
    \includegraphics[width=0.43\linewidth]{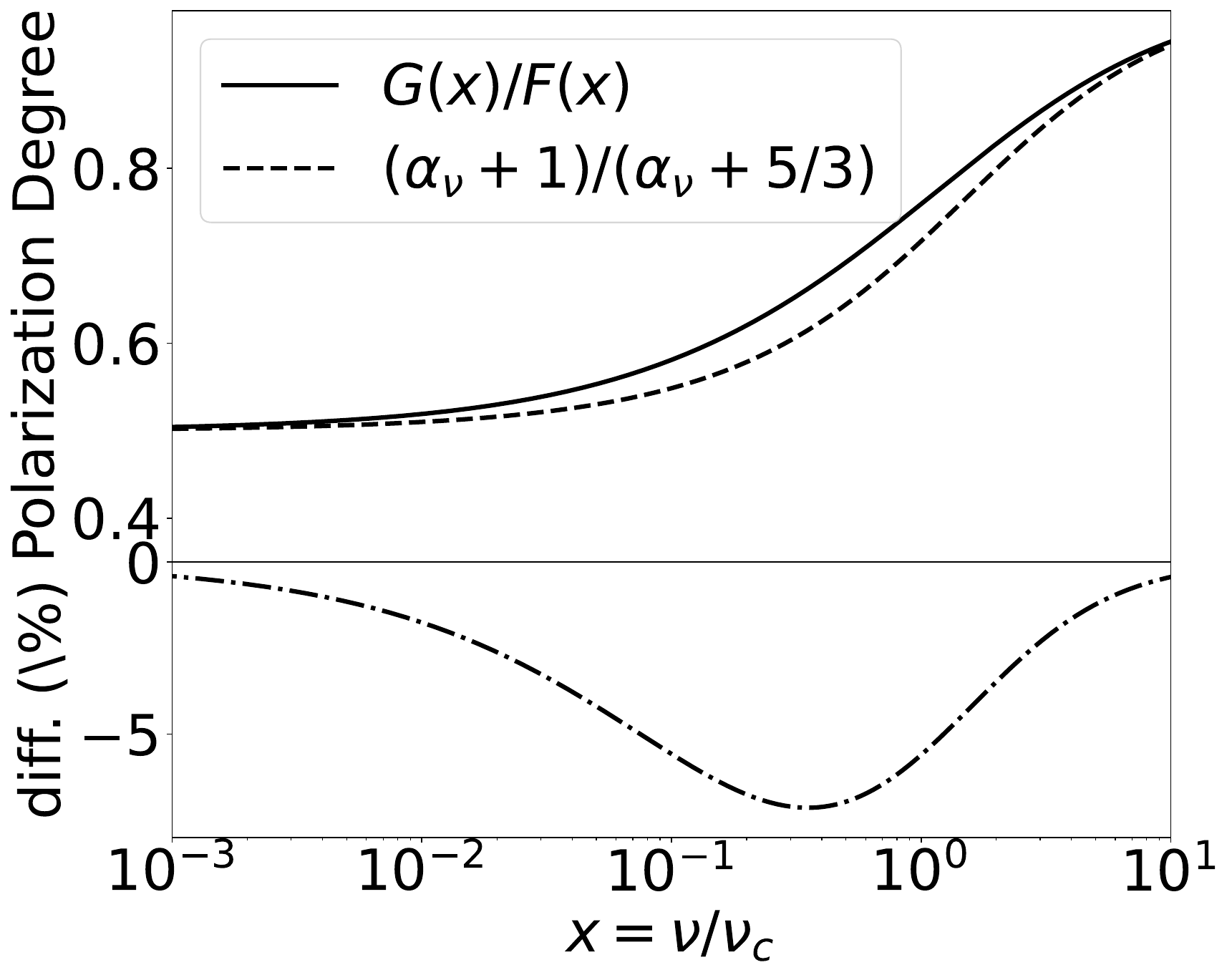}
    \caption{Synchrotron emission of isotropic mono-energy electrons.
    \emph{Left}: total flux (solid line) and polarised flux (dashed line) of the synchrotron emission with arbitrary normalisation in flux. 
    \emph{Right}: polarisation degree calculated analytically versus using 
    the generalised PD formula. 
    The percentage difference of these 
    two methods is always smaller than 7\%.}
    \label{fig:mono-energy}
\end{figure*}

For isotropic mono-energy particle distribution, the total intensity is proportional to $F(x)$ and the polarised intensity is proportional to $G(x)$,
where $x = \nu/\nu_c$.
The PD is therefore $= G(x)/F(x)$.
Figure~\ref{fig:mono-energy} shows
    the total and polarised synchrotron spectrum, and the PD calculated using two methods.
It shows that the generalised formula
    PD $= (\alpha_\nu+1)/(\alpha_\nu + 5/3)$
    works well even for the mono-energy
    case.
This is surprising given that the formula
    is originally derived from an 
    infinitely extended power law.
It is unclear the physical or mathematical
    reasons behind the validity of this formula.

\begin{figure*}
    \centering
    \includegraphics[width=0.85\linewidth]{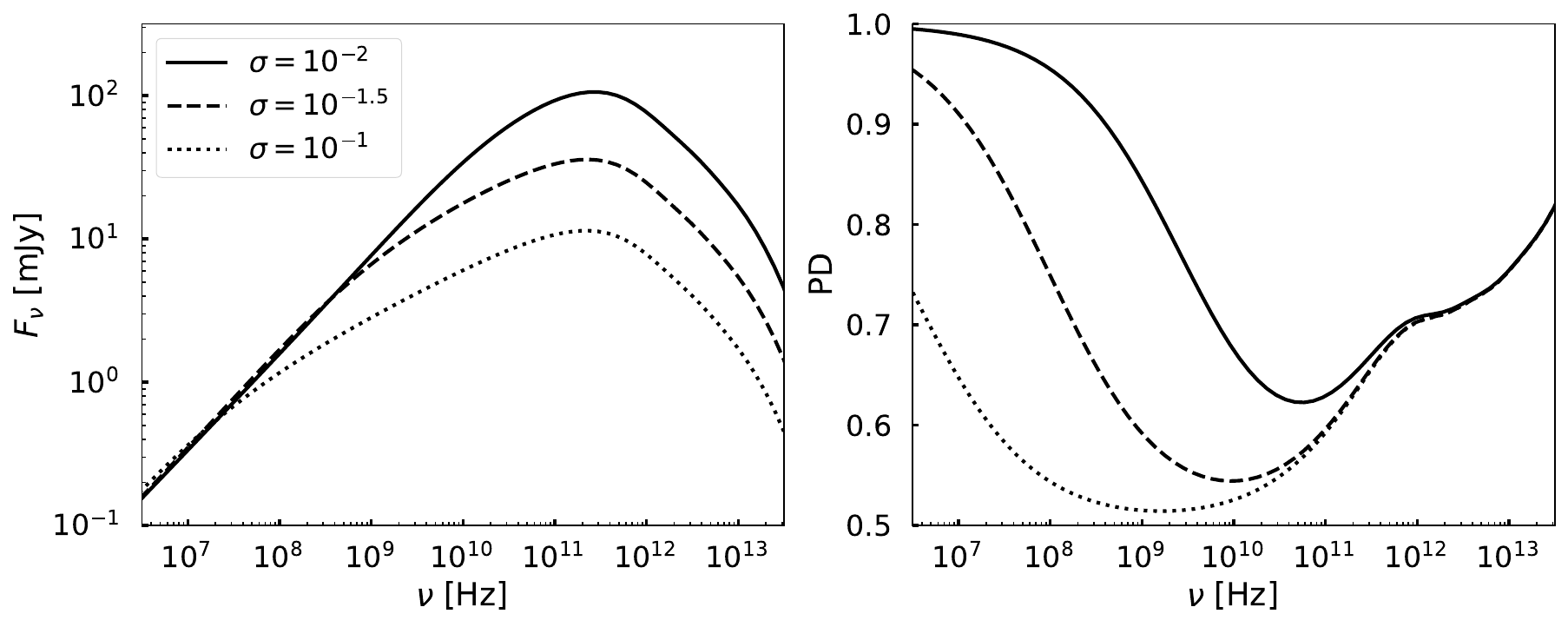}
    \caption{Reproducing Figure 5 in 
    \citet{Yang2018ApJ} and extending the calculation to polarisation. 
    The model parameters are:
    the power-law lower cut-off $\gamma_1 = 10^2$,
    the power-law upper cut-off $\gamma_2 = 10^3$,
    the power-law index $p = 2$,
    the pitch-angle distribution model parameters
    $\sigma = 10^{-1}, 10^{-1.5}, 10^{-2}$ and
    $\tilde{\alpha}_0 = \pi/4$, 
    the viewing angle $\theta = \pi/4$,
    $B = 1$\,G,
    $\delta_D = 10$,
    $N_{e, 0} = 10^{48}$, 
    and $D = 1$\,Gpc.}
    \label{fig:yang}
\end{figure*}

\section{Calculating the polarised emission using models in Yang \& Zhang (2018)} \label{app:yang_zhang}

In \cite{Yang2018ApJ}, they studied how anisotropic distribution of electrons changes the total power of the synchrotron spectrum. 
One purpose of our work is to also examine the impact on the polarised emission for electrons with anisotropic distribution.
We think it would be good idea to replicate their models with 
    polarisation calculations included.

There are a couple of differences between their work 
    and our work. 
One is that they have an additional $1/\sin^2\tilde{\alpha}$ factor
    in Eq.~\ref{eq:W_perp} and \ref{eq:W_para} to account for the difference in received and emitted power
    \citep{Rybicki1979rpa}.
As we discussed in the main text, whether to include this
    factor depends on the astrophysical context.
The inclusion of $1/\sin^2\tilde{\alpha}$ implicitly indicates that
    they considered a uniform magnetic field with electrons
    moving in perfect helical motion within the magnetic field.
In addition, they considered 
    electrons with a bulk motion,
    parametrised by the Doppler factor $\delta_D$.
We herein reproduce their model shown 
    in Figure 5 in \cite{Yang2018ApJ},
    in which electrons follow a power-law distribution with
    anisotropic distribution modelled as
\begin{equation}
    g_e(\tilde{\alpha}) =
    \frac{1}{\sqrt{2\pi\sigma^2}}
    \exp\left( 
    -\frac{(\sin\tilde{\alpha} - \sin\tilde{\alpha}_0)^2}{2\sigma^2} 
    \right)  \ . 
\end{equation}
The particle energy distribution follows
    a power law with index $p=2$,
    lower cut-off $\gamma_1 = 10^2$,
    and higher cut-off $\gamma_2 = 10^3$.
Figure \ref{fig:yang} is calculated following the recipe in \cite{Yang2018ApJ}.
Figure~\ref{fig:yang} shows the total power and the PD of the synchrotron
    radiation using their model. 
The left panel is the same as Figure 5 in \cite{Yang2018ApJ}.
The results show that, even if the electrons follow
    a power-law distribution, the anisotropy in momentum distribution
    causes very different total power spectrum 
     and polarisation property. \\

%


\bsp	
\label{lastpage}
\end{document}